\documentclass[article,10pt]{article}
\usepackage{amsmath}
\usepackage{amsfonts}
\usepackage{array}
\usepackage{enumerate}
\usepackage{graphicx}
\usepackage{centernot}
\usepackage{amssymb}
\usepackage{amsthm}
\usepackage[version=4]{mhchem}
\usepackage{xcolor}
\usepackage{chemarr}
\usepackage{mathtools}
\usepackage[margin=1in]{geometry}

\newtheorem{proposition}{Proposition}
\newtheorem{lemma}{Lemma}

\newtheorem{remark}{Remark}

\newcommand{\norm}[2][\relax]{\ifx#1\relax \ensuremath{\left\Vert#2\right\Vert} \else \ensuremath{\left\Vert#2\right\Vert_{#1}}\fi}

\begin{document}

\title{On the anti-quasi-steady-state conditions of enzyme kinetics} 

\author{Justin Eilertsen\\
            Mathematical Reviews\\ American Mathematical Society\\
            416 4th Street\\Ann Arbor, MI, 48103\\
            e-mail: {\tt jse@ams.org}\\\\
        Santiago Schnell\\
            Department of Biological Sciences and\\
            Department of Applied and Computational Mathematics and Statistics\\
            University of Notre Dame\\
            Notre Dame, IN 46556\\
            e-mail: {\tt santiago.schnell@nd.edu}\\\\
        Sebastian Walcher\\
            Mathematik A, RWTH Aachen\\
            D-52056 Aachen, Germany\\
            e-mail: {\tt walcher@matha.rwth-aachen.de}}


\maketitle

\begin{abstract} {Quasi-steady state reductions for the irreversible Michaelis--Menten 
reaction mechanism are of interest both from a theoretical and an experimental design perspective. 
A number of publications have been devoted to extending the parameter range where reduction is possible,
via improved sufficient conditions. In the present note, we complement these results by 
exhibiting local conditions that preclude quasi-steady-state reductions (anti-quasi-steady-state), in 
the classical as well as in a broader sense. To this end, one needs to obtain necessary (as opposed to 
sufficient) conditions and determine parameter regions where these do not hold. In particular, we 
explicitly describe parameter regions where no quasi-steady-state reduction (in any sense) is applicable 
(anti-quasi-steady-state conditions), and we also show that -- in a well defined sense -- these parameter 
regions are small. From another perspective, we obtain local conditions for the accuracy of standard or 
total quasi-steady-state. Perhaps surprisingly, our conditions do not involve initial substrate.}


\end{abstract}

\section{Introduction} 
In the thermodynamic limit, the temporal behavior of a chemical reaction is accurately modelled by a 
system of nonlinear ordinary differential equations that depend on a set of parameters. Furthermore, 
it is often possible to reduce the number of equations that delineate a specific reaction, especially 
in cases where the reaction contains multiple, disparate timescales. In enzyme kinetics, a particular 
class of model reductions arising from disparities in timescales is the class of quasi-steady-state 
(QSS) reductions. As an important approximation method, QSS reductions are extensively used in
applied mathematics, especially in the study of biochemical kinetics in mathematical biology.

In the classical sense, QSS reductions are justifiable when the concentrations of certain reactants 
change very slowly with respect to other species in the reaction. For example, the enzyme-substrate 
intermediate complex can be a short-lived chemical species that can react much faster than the substrate 
over the course of the chemical reaction. A large body of work in the literature has been devoted to 
identifying parameter regions that allow QSS reductions, from various perspectives. The majority of 
these studies purport that the QSS reduction scenario is common in enzyme catalyzed reactions. However, 
a pertinent question has garnered little attention: what does ``common" mean in a quantitative sense 
or, more precisely, how does one demarcate the region in parameter space where QSS reduction is
\textit{unfavorable}? In essence, what are the anti-quasi-steady-state (anti-QSS) conditions? 
In order to avoid potential misconceptions, we restate the question in a more precise
manner: We do {\em not} ask for parameter conditions under which existing sufficient criteria for 
(some type of) QSS fail to hold, but we seek to identify parameter conditions under which the dynamics 
does not reduce to a one dimensional invariant manifold, resp.\ where approximation due to a prescribed 
QSS assumption is inaccurate. From a mathematical perspective, we thus do not search for sufficient 
conditions for QSS, but we search -- via the contrapositive -- for necessary QSS conditions.

In this work, we consider the familiar irreversible Michaelis--Menten reaction mechanism
\begin{equation}\label{mm1}
    \ce{$S$ + $E$ <=>[$k_1$][$k_{-1}$] $C$ ->[$k_2$] $E$ + $P$ }.
\end{equation}
The ordinary differential equations that describe the time courses of the substrate, $s$, enzyme, 
$e$, and substrate-enzyme intermediate complex, $c$, concentrations are
\begin{equation}\label{eqmmirrev}
\begin{array}{rclclcl}
\dot s&=& -k_1e_0s&+&(k_1s+k_{-1})c & &  \\
\dot c&=& k_1e_0s&-&(k_1s+k_{-1}+k_2)c & &\\
\end{array}
\end{equation}
with parameters $e_0,s_0,k_1,k_{-1},k_2$.\footnote{$k_1$, $k_2$ and $k_{-1}$ are rate constants, 
$s_0$ is the initial substrate concentration, and $e_0$ is the initial enzyme concentrations. 
Both $s_0$ and $e_0$ correspond to conserved quantities.} 

The standard quasi-steady state assumption (sQSSA) for the Michaelis--Menten reaction mechanism
assumes that the fast chemical species, the complex, is in QSS with respect to the 
slow chemical species, the substrate. From this, one derives the familiar Michaelis--Menten equation
\begin{equation}\label{smmred}
    \dot s =-\dfrac{k_1k_2e_0s}{k_{-1}+k_2+k_1s}=-\dfrac{k_2e_0s}{K_{\rm M}+s}
\end{equation}
with the Michaelis constant $K_{\rm M}=(k_{-1}+k_2)/k_1$.
Beyond QSS for certain chemical species, the name QSS is also commonly used for more general types 
of reduction, in particular those induced by singular perturbations. We will adopt this broad usage 
here, but in our analysis we will distinguish between reductions stemming from a particular QSSA 
(such as standard QSS for complex) and general singular perturbation reductions, from
a slow-fast timescale separation. A list of common scenarios for timescale separation and reductions 
is given in Patsatzis and Goussis~\cite{PaGo}. We are interested in reductions that hold {\em over 
the whole course of the reaction} (possibly after a short initial phase), and put less emphasis on intermediate behavior and approximations. In particular, we focus on the behavior near the stationary point, 
which is relevant for the long-time evolution of solutions. In this work, we do 
not explore the implications of our analysis to the inverse problem (measurements and parameter 
estimation). In general, the validity of the QSS, while necessary, does not alone ensure the 
well-posedness of the inverse problem~\cite{Strob}. Additional analysis would be required before 
anything definitive can be said about the inverse problem.

As our vantage point, we state the minimal requirement for validity of QSS (or singular perturbation) 
reductions of any kind. The existence of a ``nearly invariant'' manifold has been established as a 
necessary condition in Goeke et al.~\cite{gwz3} (see also the discussion in Eilertsen et al.~\cite{ERSW} 
for the open Michaelis--Menten reaction mechanism), as the basic minimal requirement. In the present paper, 
we strengthen this requirement by including attractivity, which is natural from an applied perspective,
when dimension reduction is the goal.

Much effort has gone into finding parameter conditions that ensure QSSA, with the result that some 
version of QSSA holds in large parts of parameter space. In view of these results, the question 
arises whether there in fact exist any parameter regions where QSSA (derived from a specific QSSA 
as well as in a broad sense) is precluded and, if so, how does one go about calculating 
such regions for this anti-QSS? 

We will provide a concrete answer to these questions for the Michaelis--Menten reaction mechanism 
by determining a minimal requirement for the justification of a local QSS reduction, which is also 
necessary for the validity of any global QSS reduction. This minimal requirement generates criteria 
for the anti-QSS, which we can use to decide when (any type of) QSS reduction is unreasonable. We 
proceed to show that (i) parameter regions where QSS (in a broad sense) fails do exist (anti-QSS), 
(ii) how to calculate such regions, and (iii) that, in a well-defined sense, they are rather small. 
Complementing these general results, we determine local obstructions to the validity of reductions 
that stem from a specific QSS assumption (such as sQSSA or tQSSA).

In this work, we will discuss two types of local obstruction to QSSA. First, there may not exist 
a particularly pronounced (concerning attractivity) local invariant manifold near the stationary 
point, and the eigenvalue ratio of the linearization at the stationary point will be relevant in 
the discussion. Second, a given candidate for a QSS manifold (based on an a priori QSS assumption) 
may violate a certain tangency condition at the stationary point, and thus yield incorrect 
estimates for the long-term behavior. For both types of obstructions, we obtain quantifiable 
criteria in terms of the reaction parameters.

In a short final section, we reconsider the most commonly used  parameter combinations from the 
literature --- Heineken et al.~\cite{hta}, and Segel and Slemrod~\cite{SSl} --- to identify 
QSS, and show that they provide very good approximations when substrate concentration is sufficiently 
high, but they may yield less reliable approximations when substrate concentration is low. 

Our results will be found of some practical utility to determine conditions precluding the use
of QSS reductions to model enzyme catalyzed reactions. It might also have major 
implications in the estimation of enzyme kinetic parameters with reduced equations via inverse 
modeling~\cite{Strob}, which will need to be explored further.

\section{A review of the literature}
The earliest \textit{mathematical} justification of the sQSSA in the Michaelis--Menten reaction
mechanism was based on singular perturbation reduction, in particular on Tikhonov's theorem (see 
Tikhonov~\cite{tikh}, Fenichel~\cite{fenichel} and Chapter 8 of the monograph \cite{verhulst} by 
Verhulst). Although originally introduced by Briggs and Haldane~\cite{BH}, 
Heineken et al.~\cite{hta} were the first to provide -- by scaling and 
non-dimensionalizing the mass action equations -- mathematical justification for the parameter
\[
\varepsilon_{HTA}:= \dfrac{ e_0}{s_0},
\]
and demonstrated via Tikhonov's theorem that the complex concentration, $c$, 
will assume a QSS whenever $\varepsilon_{HTA}\ll 1$. It should be noted that boundedness of $s_0$ 
is implicitly required in their work, since Tikhonov's theorem is not suited for a blow-up of 
the domain of definition.

Typical numerical simulations appear to confirm the sufficiency of the qualifier 
$\varepsilon_{HTA}\ll 1$. The notion of QSS in a singular perturbation context requires the 
existence of fast (short) and slow (long) timescales, and indeed $\varepsilon_{HTA}=k_1 e_0/k_1 s_0$ 
is a ratio of timescales (although disguised as a concentration ratio). In their seminal paper, 
Segel and Slemrod~\cite{SSl} (continuing Segel~\cite{Seg}) derived the dimensionless parameter
\[
\varepsilon_{SSl}:= \dfrac{ k_1e_0}{k_{-1}+k_2+k_1s_0}
= \dfrac{ e_0}{K_{\rm M}+s_0}
\]
by timescale arguments, to ensure sQSS for complex concentration $c$, and they proved 
the existence of a sQSS (by singular perturbation) reduction as $\varepsilon_{SSl}\to 0$, assuming 
some bound on $s_0$. Extending the timescale arguments introduced in \cite{Seg,SSl},  
Borghans et al.~\cite{BBS} established the small parameter
\[
\varepsilon_{BdBS}:=\dfrac{K e_0}{(K_{\rm M}+e_0+s_0)^2}
\]
under the assumption of {\em total QSS} or tQSS, where $K=k_2/k_1$. This tQSS qualifier suggested 
that the QSS for complex was valid over a much larger parameter region than previously thought. 
Specifically, Borghans et al.~\cite{BBS}  showcased the fact that $\varepsilon_{BdBS}\ll 1$ whenever
\begin{equation*}
    \varepsilon_{SSl}\ll 1, \qquad k_2 \ll k_{-1}, \qquad \text{or} \qquad k_1/k_2 \ll e_0+s_0,
\end{equation*}
and concluded that ``{\em the tQSSA will always be at least roughly valid.}"

Modifying the approach in \cite{BBS}, Tzafriri~\cite{tza} proposed another, more intricate small 
parameter. By estimates on nonlinear timescales, Tzafriri~\cite{tza} asserted that the tQSS for 
$c$ is valid whenever
\begin{equation*}
    \varepsilon_{T}:= \cfrac{t_C}{t_S} \ll 1,
\end{equation*}
where the timescales $t_C$ and $t_S$ are given by
\begin{equation*}
t_C:=\cfrac{1}{k_1\sqrt{(K_{\rm M}+e_0+s_0)^2-4e_0s_0}}, \quad t_S:=\cfrac{2s_0}{k_2(K_{\rm M}+e_0+s_0 -\sqrt{(K_{\rm M}+e_0+s_0)^2-4e_0s_0})}.
\end{equation*}
Collectively, these seem to extend the applicability of the QSSA to an even larger region in 
parameter than the one derived by Borghans et al.~\cite{BBS}. Furthermore, with a straightforward 
calculation, Tzafriri~\cite{tza} demonstrated that 
\begin{equation*}
    \varepsilon_T \leq \cfrac{K}{4K_{\rm M}} \leq \cfrac{1}{4},
\end{equation*}
drawing from this inequality the same conclusion as Borghans et al.~\cite{BBS}: ``{\em QSS for complex 
is roughly always valid}'', due to $\varepsilon_T<1$. From a mathematical perspective, this argument 
is problematic, and seems to be based on a too literal interpretation of the condition 
``$\varepsilon_T\ll 1$''.

More recently, careful analytical and numerical work by Patsatzis and Goussis~\cite{PaGo} showed 
that QSSA (in some suitable version, characterized by timescale disparity) is valid for a wide 
range of parameters. Qualitatively, if not quantitatively, this further supports the claim of 
Borghans et al.~\cite{BBS}, and Tzafriri~\cite{tza}.

The aforementioned parameters were all derived from \textit{physical} timescale estimates, and 
the mathematical legitimacy of the QSSA was attributed (directly or indirectly) to singular 
perturbation theory. Interestingly, however, only Patsatzis and Goussis~\cite{PaGo} sought to 
estimate timescales via the stiffness ratio of the Jacobian, despite the fact that eigenvalue 
disparity is fundamental for the application of singular perturbation reduction. For a general 
perspective on eigenvalue disparity and invariant sets, we recommend the readers to consult
Chicone \cite{chicone}, Theorem 4.1.  It should be noted that Patsatzis and Goussis~\cite{PaGo} 
were not the first to introduce eigenvalue methods. The parameter
\[
\varepsilon_{RS}:= \dfrac{e_0}{K_{\rm M}},
\]
which was introduced by Reich and Selkov~\cite{ReSe} and justified by Palsson and Lightfoot~\cite{PaLi}, 
characterizes the local situation (being determined by the eigenvalue ratio of the Jacobian) 
at the stationary point. Note that the qualifier $\varepsilon_{RS}\ll 1$ is far more restrictive 
than those of Segel and Slemrod~\cite{SSl}, Borghans et al.~\cite{BBS}, and Tzafriri~\cite{tza}.

While the corpus of QSS analysis suggests that the QSSA is valid over a large region in parameter 
space (especially the tQSSA), it is unclear as to \textit{where} in parameter space the QSSA is 
invalid (anti-QSSA). To fill this gap, we will turn matters around and identify parameter ranges 
where QSS is precluded. Specifically, in what follows, we ask: Where precisely is the QSS invalid, 
and ``how bad can things actually get?" While most of our results are stated for the 
Michaelis--Menten reaction mechanism, they may be generalized to any planar system that admits an attracting node. 

\section{Local conditions for the QSSA}\label{loccsec}
Before we go about determining parameter regions where QSS reduction is invalid, we first will 
identify minimal {\em local} requirements for validity of QSS reductions. We start from the 
crucial insight that any version of QSSA requires the existence of a distinguished 
``nearly invariant'' manifold:
\begin{itemize}
    \item A QSSA for some chemical species (such as complex in Michaelis--Menten reaction mechanism) 
    in some parameter region defines a manifold $Y$ in phase space via setting their rates of change 
    equal to zero (e.g.\ $\dot c=0$ in Michaelis--Menten reaction mechanism). This distinguished 
    manifold will be called the {\em QSS manifold} in the following. While one does not require 
    invariance of $Y$ for the system, validity of QSS means that there must exist an invariant 
    manifold close to $Y$; see Goeke et al.~\cite{gwz3} for more background. We add a stronger 
    requirement here, which ensures relevance for reduction: This nearly invariant manifold 
    should also attract all nearby trajectories. 
    \item Interpreting QSS as a singular perturbation scenario, one also has a distinguished 
    manifold, viz.\ the critical manifold, and Fenichel provides conditions for an invariant 
    manifold to persist for all sufficiently small perturbations.
\end{itemize}
So, in both scenarios a distinguished manifold exists. This fact was clearly observed by Schauer 
and Heinrich~\cite{SchaHe} in their derivation of QSSA conditions. Segel and Slemrod \cite{SSl} 
also work implicitly with the assumption of a distinguished invariant manifold when they base a 
slow timescale estimate from the Michaelis-Menten equation \eqref{smmred} (which requires 
near-invariance of the QSS manifold; see Goeke et al.~\cite{gwz3}), and this is also the 
basis of the arguments given in Borghans et al.\ \cite{BBS}, as well as Tzafriri~\cite{tza}. 
Please note that sQSSA and tQSSA share the same QSS manifold.

Since all solutions of the Michaelis--Menten reaction mechanism converge to the stationary 
point $0$, this distinguished manifold should correspond to a distinguished manifold at 
the stationary point, which we call a {\em local QSS manifold}. Local considerations of 
QSS exist in the literature. Implicitly, Palsson and Lightfoot~\cite{PaLi} dealt with 
local invariant manifolds, while Roussel and Fraser~\cite{RouFra} and Schnell and Maini~\cite{ScMa} 
obtained an iterative  scheme to approximate a global invariant manifold starting from a 
local one.

\subsection{Nodes and their properties} 
Recall that a stationary point of a planar differential equation is called a node if both eigenvalues 
of its linearization are real and have the same sign. Here we are interested in attracting nodes, 
with the eigenvalues satisfying $\lambda_2\leq \lambda_1<0$. For the readers' 
convenience, we give a brief account regarding eigenvalue ratios and dynamics near a node.
\begin{itemize}
    \item 
If the eigenvalues $\lambda_1,\,\lambda_2$ are different,  then, in some neighborhood of the 
stationary point, all but at most two nonconstant trajectories approach the stationary point 
tangent to the eigenspace for $\lambda_1$, possibly with the remaining two approaching tangent 
to the eigenspace of $\lambda_2$. Please cosult Perko~\cite{Per}, Section 2.10; in particular 
Theorem~4  for details.\footnote{ Whenever $\lambda_2$ is not an integer multiple of $\lambda_1$ then 
there are two exceptional trajectories. In the remaining cases there may be two or none; 
see Walcher~\cite{WPoinc}, Theorem 2.3.} So the natural candidate for a local QSS 
manifold is the eigenspace for the ``slow'' eigenvalue $\lambda_1$. If a (global) QSS manifold 
$Y$ is given then one can test the accuracy of approximation by the QSS reduction by comparing 
its tangent space at $0$ with the eigenspace. In Section~\ref{ltsubsec}, we will discuss this 
matter in detail, and show quantitatively that a deviation in tangent directions implies 
erroneous timescale estimates in the long term.
\item From a general perspective, the observation regarding distinguished local invariant 
manifolds implies that no type of QSS may be possible. The argument is as follows:
\begin{itemize}
    \item Whenever $\lambda_1\not=\lambda_2$ then a local QSS manifold exists, viz.\ the 
    eigenspace for $\lambda_1$. But there remains to quantify attractivity.  
    \item Consider first the uncoupled linear system
\[
\begin{array}{rcl}
\dot x_1&=&\lambda_1x_1\\
\dot x_2&=&\lambda_2x_2\\
\end{array}
\text{ with solution }
\begin{array}{rcl}
x_1&=& u_1\exp(\lambda_1 t)\\
x_2&=& u_2\exp(\lambda_2 t)\\
\end{array}.
\]
The flow of this linear system is contracting, but the contraction is stronger in $x_2$-direction, 
as seen from
\[
\dfrac{x_2}{x_1}=\dfrac{u_2}{u_1}\exp(\lambda_2-\lambda_1)t.
\]
Substitution of the slow local timescale $|\lambda_1|^{-1}$ for $t$ yields an intrinsic 
contraction ratio
\[
\exp\left(1-\dfrac{\lambda_2}{\lambda_1}\right),
\]
which characterizes the attractivity of the $x_1$-axis. 
\item Now for any linear system with different eigenvalues one obtains an analogous result 
with the eigenspaces for $\lambda_1$ and $\lambda_2$, and this carries over locally to nonlinear 
systems, via an analytic transformation to Poincar\'e-Dulac normal form (see e.g.\ 
Bruno \cite{Bru})\footnote{For a node the eigenvalues $\lambda_i$ lie in a Poincar\'e domain. 
We do not discuss here the technicality appearing when $\lambda_2$ is an integer multiple 
of $\lambda_1$; the statement about the contraction ratio remains unchanged.}.
Thus, if the eigenvalue ratio $\lambda_1/\lambda_2$ is small then, near the stationary point, 
trajectories will be strongly attracted to the eigenspace for $\lambda_1$.
\item On the other hand,  for  $\lambda_1/\lambda_2\approx 1$ the order of tangency will 
be near zero, and from a quantitative perspective there is no distinct QSS property. 
\item In the borderline case when both eigenvalues are equal, but the linearization is 
not semisimple (which is the case for Michaelis--Menten reaction mechanism), solutions approach 
the stationary point with a single tangent direction, but before doing so they make sharp turns 
in arbitrarily small neighborhoods of the stationary point. The flow of the linearization 
is a shear mapping, with contraction ratio $1$.
\end{itemize}
\end{itemize}

\subsection{The local attractivity condition for QSSA} 
By the arguments above, an appropriate minimal condition is of the form
\begin{equation}\label{mincond}
\dfrac{\lambda_1}{\lambda_2}\leq {\rho<1},
\end{equation}
for some {$\rho$}, with small {$\rho$} corresponding to strong contraction. Here the choice 
of {$\rho$} is up to deliberation. Inspection indicates that one definitely should require 
{$\rho\leq 1/2$}, and {$\rho\leq 1/4$} might be more appropriate; 
{see {{\sc Figure}}~\ref{fig:2}.}\footnote{{The choice $\rho=1/4$ is suggested by the fact that $\lambda_1/\lambda_2{\approx} 1/4$ is still 
deemed acceptable in Borghans et al.~\cite{BBS}, see their Figure 2(c), with $e_0=1$. In any case, the desired degree of accuracy determines $\rho$.}}

\begin{remark}
In general, $\rho \to 0$ does not automatically certify the accuracy of the 
sQSSA~(\ref{smmred}). However, it does qualify local QSS; the specific reduction that holds 
(i.e., the sQSSA, tQSSA, \ldots, etc. or any from the list in Patsatzis and Goussis~\cite{PaGo}) 
depends on which parameters are \textit{comparatively} small.
\end{remark}

\medskip
We rephrase the minimal condition, avoiding square roots. Let $A$ be a real $2\times 2$ matrix 
with (nonzero) real eigenvalues $\lambda_1,\,\lambda_2$ of equal sign. Then the coefficients 
of the characteristic polynomial $\tau^2+\sigma_1\tau+\sigma_2$ of $A$ satisfy
\[
\sigma_1=-(\lambda_1+\lambda_2),\quad \sigma_2=\lambda_1\lambda_2.
\]

We have the identity
\begin{equation}\label{lamsig2}
\frac{\lambda_2}{\lambda_1} + \frac{\lambda_1}{\lambda_2}=\frac{\lambda_1^2+\lambda_2^2}{\lambda_1\lambda_2}=\frac{\sigma_1^2-2\sigma_2}{\sigma_2}.
\end{equation}
Moreover, for positive $x$, the function defined on the interval $(0,1]$ by $x\mapsto x+\frac1x$ 
is strictly decreasing and attains its minimum $2$ at $x=1$. We thus obtain the following Lemma.
\begin{lemma}\label{elemlem}
For $0<\delta\leq 1$ one has
\begin{equation}
\frac{\lambda_1}{\lambda_2}\geq \delta \Rightarrow \frac{\sigma_1^2-2\sigma_2}{\sigma_2}\leq \delta+\frac1\delta\Rightarrow \frac{\sigma_2}{\sigma_1^2-2\sigma_2}\geq \frac{\delta}{1+\delta^2},
\end{equation}
as well as
\begin{equation}
\frac{\lambda_1}{\lambda_2}\leq \delta \Rightarrow \frac{\sigma_1^2-2\sigma_2}{\sigma_2}\geq \delta+\frac1\delta\Rightarrow \frac{\sigma_2}{\sigma_1^2-2\sigma_2}\leq \frac{\delta}{1+\delta^2}.
\end{equation}
In particular
\begin{equation}
\lambda_1= \lambda_2 \Leftrightarrow\frac{\sigma_1^2-2\sigma_2}{\sigma_2}= 2.
\end{equation}
\end{lemma}
\noindent Consequently, we see that the \textit{minimal} requirement for the QSSA is a sufficiently 
small $\delta$; see {{\sc Figure}}~\ref{fig:2}.

\subsection{Application to Michaelis--Menten reaction mechanism}
We now turn to the Michaelis--Menten reaction mechanism~(\ref{mm1}), and characterize parameter 
regions where local QSS conditions (QSS being understood in a broad sense) near the stationary 
point do or do not hold. Note that non-validity of local conditions implies global non-validity 
of QSSA (anti-QSSA).

Specializing, we find from Lemma \ref{elemlem} and elementary computations:
\begin{proposition}\label{prop1}
Let $0<\delta\leq1$. Then 
\[
\frac{\lambda_1}{\lambda_2}\geq \delta \Leftrightarrow k_1k_2e_0\geq \frac{\delta}{1+\delta^2}\left( (k_{-1}+k_2+k_1e_0)^2-2k_1k_2e_0\right)
\]
as well as (equivalently) 
\[
\frac{\lambda_1}{\lambda_2}\leq \delta \Leftrightarrow k_1k_2e_0\leq {\frac{\delta}{1+\delta^2}}\left((k_{-1}+k_2+k_1e_0)^2-2k_1k_2e_0\right).
\]
In particular the expression $\cfrac{\sigma_1^2-2\sigma_2}{\sigma_2}$ attains its minimum 2 
(with $\lambda_1=\lambda_2$) if and only if
\[
k_1e_0=k_2 \quad \text{and} \quad k_{-1}=0.
\]
\end{proposition}

\begin{remark}
Informally, one may state a consequence of the Proposition as
\begin{equation}\label{lamdel}
    \dfrac{\lambda_1}{\lambda_2}\ll1\Longleftrightarrow \dfrac{K e_0}{(K_{\rm M}+e_0)^2}\ll1.
\end{equation}
It is instructive to compare the right hand side of \eqref{lamdel} to the parameter 
$\dfrac{K e_0}{(K_{\rm M}+e_0+s_0)^2} $ due to Borghans et al.\ \cite{BBS}. The expressions are 
similar, but in the local condition the initial substrate concentration does not appear. 
Moreover, one sees that 
\[
\dfrac{K e_0}{(K_{\rm M}+e_0)^2}\leq \varepsilon_{RS},
\]
thus smallness of the Reich-Selkov parameter always ensures a large eigenvalue gap.
\end{remark}

We now show that {\em there exist local obstructions to QSSA}, hence the summary assertions 
by Borghans et al.~\cite{BBS}, and Tzafriri~\cite{tza} that tQSSA is ``roughly valid'' for all 
parameter combinations are not sustainable.

\begin{figure}
  \centering
    \includegraphics[width=8.0cm]{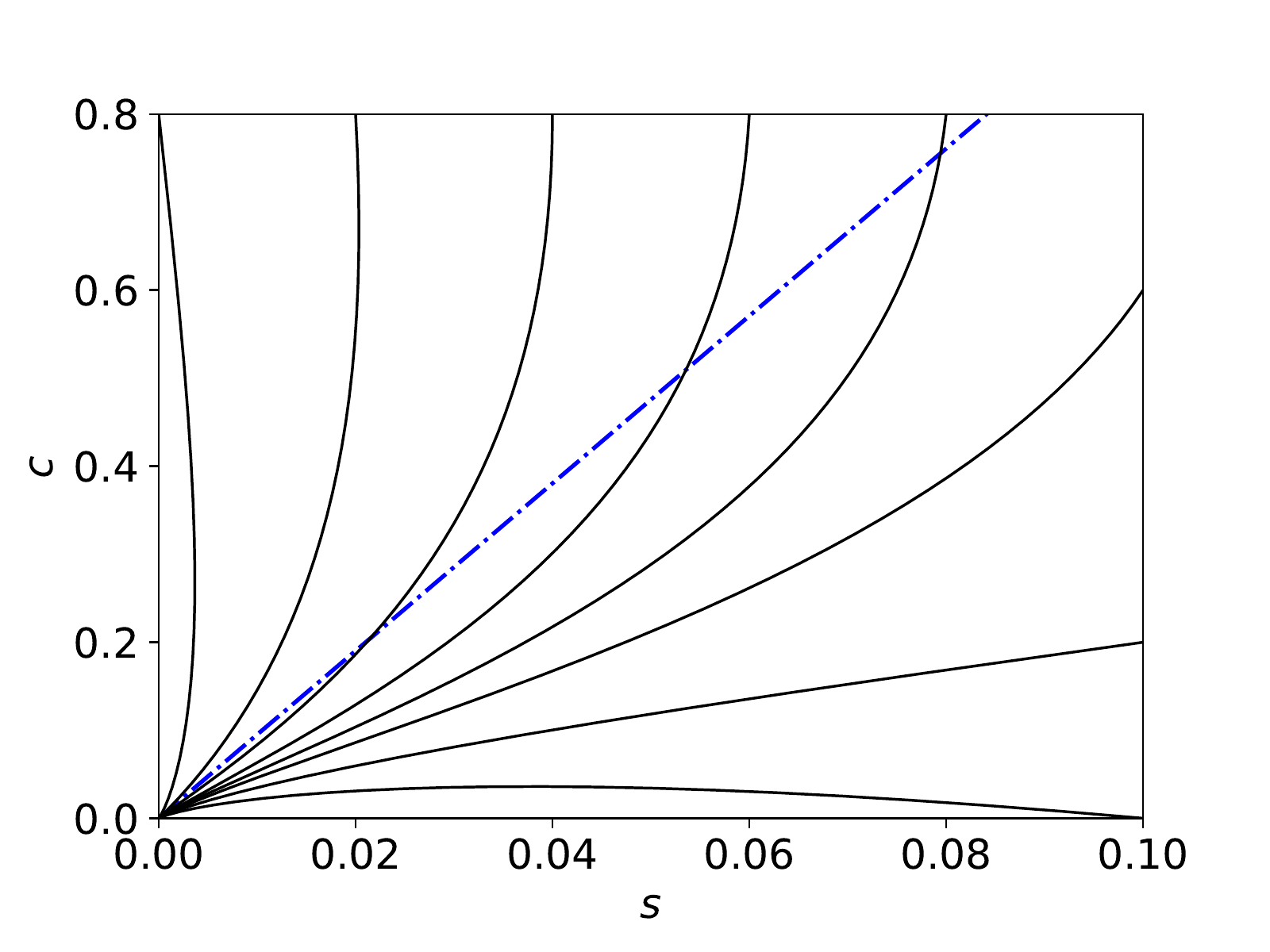}
    \includegraphics[width=8.0cm]{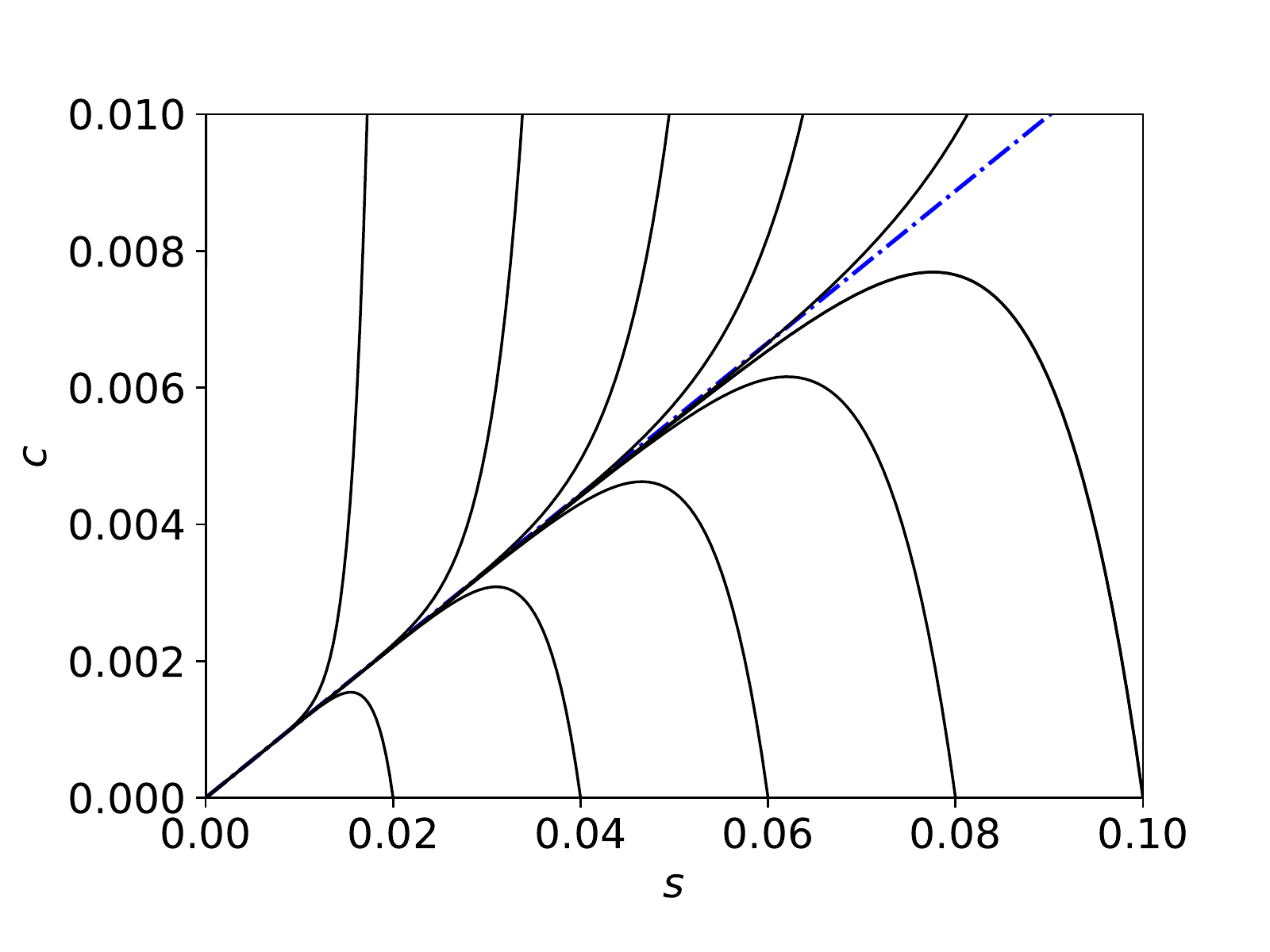}
  \caption{{\textbf{The existence of a sufficiently large spectral gap (thus sufficiently 
  small $\delta$) constitutes the minimal requirement for the validity of a QSS}. In both panels, 
  the thin black curves are numerical solutions to the mass action equations~(\ref{eqmmirrev})
  for the Michaelis--Menten reaction mechanism.   The dashed/dotted blue line is the slow eigenspace. 
  {{\sc Left}}: Parameter values: $k_1=1.0$, $k_2=1.0$, $e_0=1.0$ and $k_{-1}=0.01$ and 
  $\delta {\approx} 1.0$. Eventually, trajectories approach   the origin in a 
  direction that is tangent to the slow eigenspace. However, since the spectral   gap is small, 
  tangency manifests very late in the reaction and is so local that it is of little use for any 
  meaningful reduction. {{\sc Right}}: Parameter values: $k_1=1.0$, $k_2=10.0$, $e_0=1.0$ and 
  $k_{-1}=0.01$. In this case $\delta{\approx} 0.1$, and trajectories 
  \textit{quickly} approach the origin in the direction of the slow eigenspace.}}\label{fig:2}
\end{figure}

Collectively, Lemma \ref{elemlem}, Proposition \ref{prop1} and the numerical simulations 
presented in the {{\sc left}} panel of {{\sc figure}} \ref{fig:2} illustrate the important 
fact that an upper bound on Segel's \cite{Seg} nonlinear timescale ratio  (such as the upper 
bound $1/4$ in Tzafriri~\cite{tza}) does not imply that the eigenvalues are disparate. Hence,
\textit{physical} timescale separation, as used in Segel and Slemrod~\cite{SSl}, Borghans 
et al.~\cite{BBS}, and Tzafrifi~\cite{tza}, is possible even in the absence of 
\textit{mathematical} timescale separation, but may fail in the accurate description of 
long-term behavior of the system dynamic. From this perspective, it must be argued that 
mathematical timescale ratios (stiffness ratios) yield a more meaningful assessment of QSSA 
legitimacy.  

On the other hand, we can substantiate, and give a precise meaning to, the following statement: 
{\em The local version of QSSA for Michaelis--Menten reaction mechanism is valid in a large 
part of the parameter range.} To this end, we ask: {How large, in a quantitative sense, is 
the region in parameter space where $\lambda_1/\lambda_2$ is close to unity?} 

\begin{figure}
  \centering
    \includegraphics[width=8.0cm]{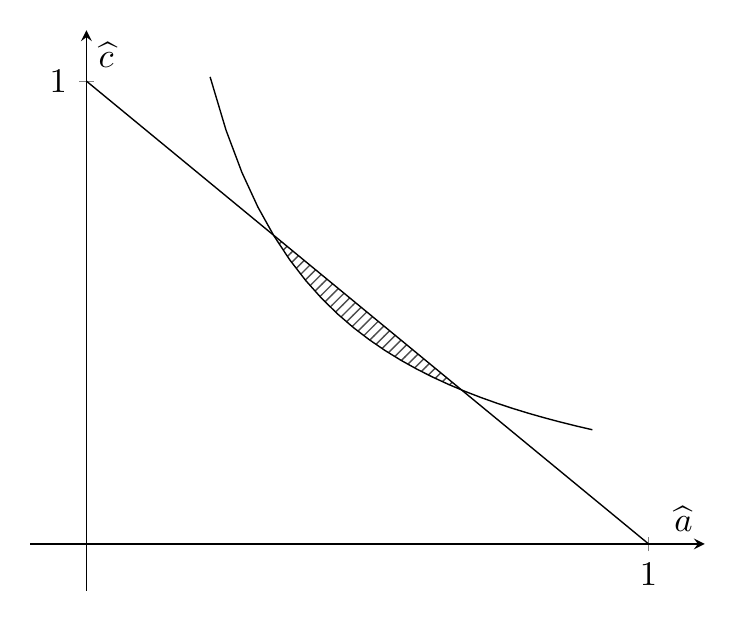}
    \includegraphics[width=8.0cm]{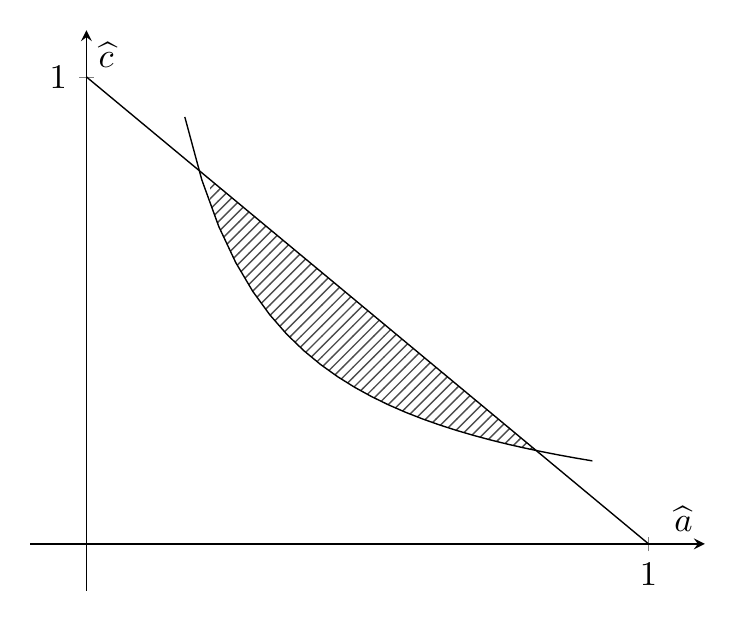}
  \caption{{{\sc Left}}: The shaded region coincides with $\delta\geq\frac12$. 
  {{\sc Right}} The shaded region coincides with $\delta\geq\frac14$. Incidentally, the midpoint 
  of the hypotenuse represents $\delta=1$. Geometrically, we see that the majority of parameter 
  combinations satisfy $\delta < 1/4$, and very few parameter combinations yield $\delta \geq 1/2.$ 
  Determining the relevant fractions of the triangle area is a matter of basic calculus. 
  One can interpret the shaded region as representing the likelihood that the eigenvalue ratio is 
  bounded below by $1/2$ ({{\sc Left}} panel) or $1/4$ ({{\sc Right}} panel) if one were to 
  randomly select a parameter ratio $(k_1e_0:k_{-1}:k_2)$ . {In real-life networks, 
  an eigenvalue ratio $\approx 1$ may be even more infrequent: These ratios occur only when 
  $k_1e_0\approx k_2$ and $k_{-1}\ll k_2$, and a meta analysis of parameter estimates from 
  experimental data may  reveal that the condition $k_{-1}\ll k_2$ is a rare circumstance.}} \label{fig:drawings}
\end{figure}

At the stationary point $0$, the Jacobian of the right-hand side equals
\[
A=\begin{pmatrix} -a & b\\ a&-(b+c)\end{pmatrix}, \quad \text{with } a=k_1e_0,\,b=k_{-1},\,c=k_2.
\]
Both eigenvalues are real and negative, with coefficients
\[
\sigma_1=a+b+c,\quad \sigma_2=ac
\]
of the characteristic polynomial.  Instead of discussing $\cfrac{\sigma_1^2-2\sigma_2}{\sigma_2}$, 
we may just as well consider its inverse
\[
\phi=\frac{\sigma_2}{\sigma_1^2-2\sigma_2}=\frac{ac}{a^2+b^2+c^2+2b(a+c)}=\frac{ac}{(a+b+c)^2-2ac}.
\]
Since this expression is homogeneous of degree zero in $a,\,b,\,c$, we may introduce normalized 
parameters
\[
\widehat a=a/(a+b+c),\quad \widehat b=b/(a+b+c),\quad \widehat c=c/(a+b+c).
\]
Then the parameter space is represented by the simplex
\[
\Delta=\{(\widehat a,\widehat b,\widehat c):\, \widehat a\geq 0,\,\widehat b\geq 0,\,\widehat c\geq 0,\,\widehat a+\widehat b+\widehat c=1\},
\]
up to scaling $a,\,b,\,c$ by the same factor.
With $\widehat b=1-\widehat a-\widehat c$ there remains to investigate
\begin{equation}\label{psieq}
\psi(\widehat a,\widehat c)=\frac{\widehat a\widehat c}{1-2\widehat a \widehat c}\;\;\text{  for  }\widehat a\geq 0,\;\;\widehat c\geq 0, \;\;\widehat a+\widehat c\leq 1.
\end{equation}
Noting
\[
\psi(\widehat a,\widehat c)\geq \gamma\Longleftrightarrow\widehat a\widehat c\geq \frac{\gamma}{1+2\gamma}\quad\text{ for }0\leq \gamma\leq 1,
\]
one may quantify and visualize eigenvalue ratios by turning to $\widehat a$ and $\widehat c$, 
and to $\psi$ in equation \eqref{psieq}. Here $\widehat a$, $\widehat c$ are confined to the triangle 
given by $\widehat a\geq 0$, $\widehat c\geq 0$, $\widehat a+\widehat c\leq 1$, and 
$\lambda_1/\lambda_2\geq \delta$ corresponds to $\psi(\widehat a,\widehat c)\geq \delta/(1+\delta^2)$. 
For $\delta$ near $1/2$, this inequality defines a rather small region in the triangle, as 
illustrated in {{\sc Figure}}~\ref{fig:drawings}.  Inspection of the {{\sc left}} panel in 
{{\sc Figure}}~\ref{fig:drawings} reveals that one might intuitively say that 
$\lambda_1/\lambda_2<\frac12$ for ``most'' parameter combinations. On the other hand, the case
$\lambda_1=\lambda_2$ once again shows that, in contrast to the assertion in Borghans et al.~\cite{BBS}, 
no kind of QSS is ``roughly valid'' over the whole parameter range; see {{\sc Figure}}~\ref{fig:2}, 
as well as {\sc{Figure}}~\ref{fig:traj} below.

\section{The local tangency condition for Michaelis--Menten reaction mechanism}\label{ltsubsec}
In the previous section, we discussed conditions for QSSA in a broad sense, with no reference to 
the particular nature of the reduction or the distinguished local manifold. In the present section, 
we start from some specific QSSA and test the quality of the approximation near the stationary point. 
Rather than deriving general estimates for local tangency conditions, we focus here on the
differential equations for the Michaelis--Menten reaction mechanism~\eqref{eqmmirrev}.

We assume that a candidate for the QSS manifold $Y$ is given in the form $c=g(s)$, with $g(0)=0$. 
This could be the manifold corresponding to standard QSSA, but also any slow invariant manifold 
from the list in Patsatzis and Goussis~\cite{PaGo}, for instance. Note that only $g^\prime(0)$ 
enters computations in the following, thus no explicit form of $g$ is needed. The eigenvalues 
of the Jacobian at $0$ are 
\[
\lambda_{1,2}=\frac12\left(-(k_{-1}+k_2+k_1e_0)\pm \sqrt{(k_{-1}+k_2+k_1e_0)^2-4k_1k_2e_0}\right),
\]
with $|\lambda_1|<|\lambda_2|$. For the sake of brevity, we disregard the case $k_{-1}=0$ here 
and in the following.

 We say that the QSS manifold $Y$ satisfies the {\em tangency condition at $0$} if the tangent vector 
 $\begin{pmatrix} 1\\ g^\prime(0)\end{pmatrix}$ is equal to the basis vector 
 $\begin{pmatrix} 1\\ \alpha\end{pmatrix}$ of the eigenspace for $\lambda_1$. Explicitly,
    \[
    \alpha=\frac1{2k_{-1}}\left(-k_{-1}-k_2+k_1e_0+\sqrt{(k_{-1}+k_2+k_1e_0)^2-4k_1k_2e_0}\right)>0,
    \]
and 
    \[
    \lambda_1=-k_1e_0+k_{-1}\alpha.
    \]
    
One will not expect the tangency condition to be satisfied exactly, but it must hold with some 
degree of accuracy to ensure global accuracy of the QSS reduction. To justify this statement, 
we will show that violation of the tangency condition, measured by
\begin{equation}\label{eq:mu}
    \mu=\left|\dfrac{k_{-1}(g^\prime(0)-\alpha)}{\lambda_1}\right|,
\end{equation}
results in incorrect long-time asymptotics for the substrate concentration.\footnote{As can be 
seen from the expression for $\mu$, the appropriate condition requires more than 
``geometric tangency''. Note in particular the denominator $\lambda_1$. But we keep the brief 
name ``tangency condition''.}
    
First, substitution of $c=g(s)$ into the first equation of \eqref{eqmmirrev} and linear 
approximation yields
    \[
    \dot s=:h(s)=h'(0)s + \text{h.o.t.}=\left(-k_1e_0+k_{-1}g^\prime(0)\right)s + \text{h.o.t.}
    \]
    On the other hand, every solution of system \eqref{eqmmirrev} approaches the stationary point 
    tangent to the eigenspace for $\lambda_1$, so $c=\alpha s$ up to higher order terms, and we 
    get the correct approximation
    \[
    \dot s=\widetilde h(s)=\left(-k_1e_0+k_{-1}\alpha\right)s + \text{h.o.t.} = \lambda_1s + \text{h.o.t.},\quad \text{whence }\widetilde h'(0)=\lambda_1.
    \]
Thus, in the course of approximately linear slow degradation of $s$, the QSSA with the manifold 
$Y$ given by $c=g(s)$, misrepresents time evolution by a factor 
\[
    \exp\left(|h^\prime(0)-\widetilde h^\prime(0)|t\right)=\exp\left(k_{-1}\left|g^\prime(0)-\alpha|\right)t\right).
    \]
    Within a characteristic time $|\lambda_1|^{-1}$ for substrate depletion at low concentrations, 
    this QSSA is incorrect by a factor 
    \begin{equation}
    \exp\left(|h^\prime(0)-\widetilde h^\prime(0)|/|\lambda_1|\right)= \exp\left(|1- h^\prime(0)/\lambda_1|\right)=\exp(\mu),
    \end{equation}
which can be quite large. 

We now specialize the discussion and ask under what circumstances is the tangency condition 
(approximately) valid for the sQSSA, as well as for tQSSA. For both these approximations, 
we have 
    \[
    h^\prime(0)=-\dfrac{k_2e_0}{K_{\rm M}}=-k_2\varepsilon_{RS}.
    \]
    Now consider the function
    \begin{equation}\label{qfunc}
    \begin{array}{rcl}
       q:& &\,[0,\infty)\to\mathbb R    \\
       x&\mapsto  & \frac12\left(-(k_{-1}+k_2+x)+ \sqrt{(k_{-1}+k_2+x)^2-4k_2x}\right)+\dfrac{K x}{K_{\rm M}}
    \end{array}
    \end{equation}
    with fixed positive parameters $k_{-1}$, k$_2$, noting $q(k_1e_0)=\lambda_1-h'(0)$. As shown 
    in the Appendix~\ref{app}, $q''>0$, hence $q$ is convex, and $q(0)=q'(0)=0$. Therefore $q$ is 
    strictly increasing with $x$ and we have $q(k_1e_0)>0$ for positive parameters. This implies 
    \begin{equation}\label{gtn}
        |h'(0)| > |\lambda_1|,\quad \mu=\dfrac{h'(0)}{\lambda_1}-1>0,
    \end{equation}
and thus the QSSA always underestimates the timescale for substrate depletion.
    
We proceed to obtain more palatable estimates for some parameter regions. Using
\[
\begin{array}{rcl}
    \widetilde h'(0)=\lambda_1&=& -k_1e_0+k_{-1}\alpha\\
      &=& \cfrac{1}{2}\left(k_{-1}+k_2+k_1e_0\right)\left(-1+\sqrt{1-\dfrac{4Ke_0}{(K_{\rm M}+e_0)^2}}\right),
    \end{array}
    \]
and the well known inequalities
    $1-x\leq\sqrt{1-x}\leq 1-x/2$ for $0\leq x\leq 1$, one obtains
    \[
    -\dfrac{2k_2e_0}{K_{\rm M}+e_0}\leq \lambda_1\leq-\dfrac{k_2e_0}{K_{\rm M}+e_0},
    \]
   equivalently:
     \begin{equation}\label{lamoneest}
    -\dfrac{K_{\rm M}+e_0}{2k_2e_0}\geq \dfrac{1}{\lambda_1}\geq-\dfrac{K_{\rm M}+e_0}{k_2e_0}.
    \end{equation}

After some elementary computations we arrive at
\[
\dfrac{e_0-K_{\rm M}}{2K_{\rm M}} \leq \dfrac{h'(0)}{\lambda_1}-1\leq \dfrac{e_0}{K_{\rm M}},\ 
\]
and thus, given (\ref{gtn}), we obtain
\[
\
\max\bigg \{\dfrac{e_0-K_{\rm M}}{2K_{\rm M}},0\bigg\} \leq \mu\leq \dfrac{e_0}{K_{\rm M}}=\varepsilon_{RS}.\ 
\]
In any case, we see
\begin{equation}\label{ineq1}
0 \leq \mu\leq \varepsilon_{RS},
\end{equation}
and therefore the tangency condition improves as $\varepsilon_{RS}\to 0$. But, if $e_0>K_{\rm M}$ so 
that $\varepsilon_{RS} > 1$, then
\begin{equation}\label{ineq2}
\dfrac12(\varepsilon_{RS}-1) \leq \mu\leq \varepsilon_{RS},
\end{equation}
and we see that the local tangency condition is violated when enzyme is abundant. Consequently, 
the timescale estimate for slow degradation of substrate is incorrect.\footnote{In fact, it is 
clear from (\ref{ineq2}) that the tangency condition is severely violated when 
$1\ll \varepsilon_{RS}$, even though $\delta$ may in fact be much less than $1$ in such cases.} 
Numerical simulations illustrate this fact; see {{\sc Figure}}~\ref{fig:tan}. In the 
Appendix~\ref{app}, we will provide more detailed estimates, distinguishing various cases for
the reaction parameters.

\begin{figure}
  \centering
    \includegraphics[width=8.0cm]{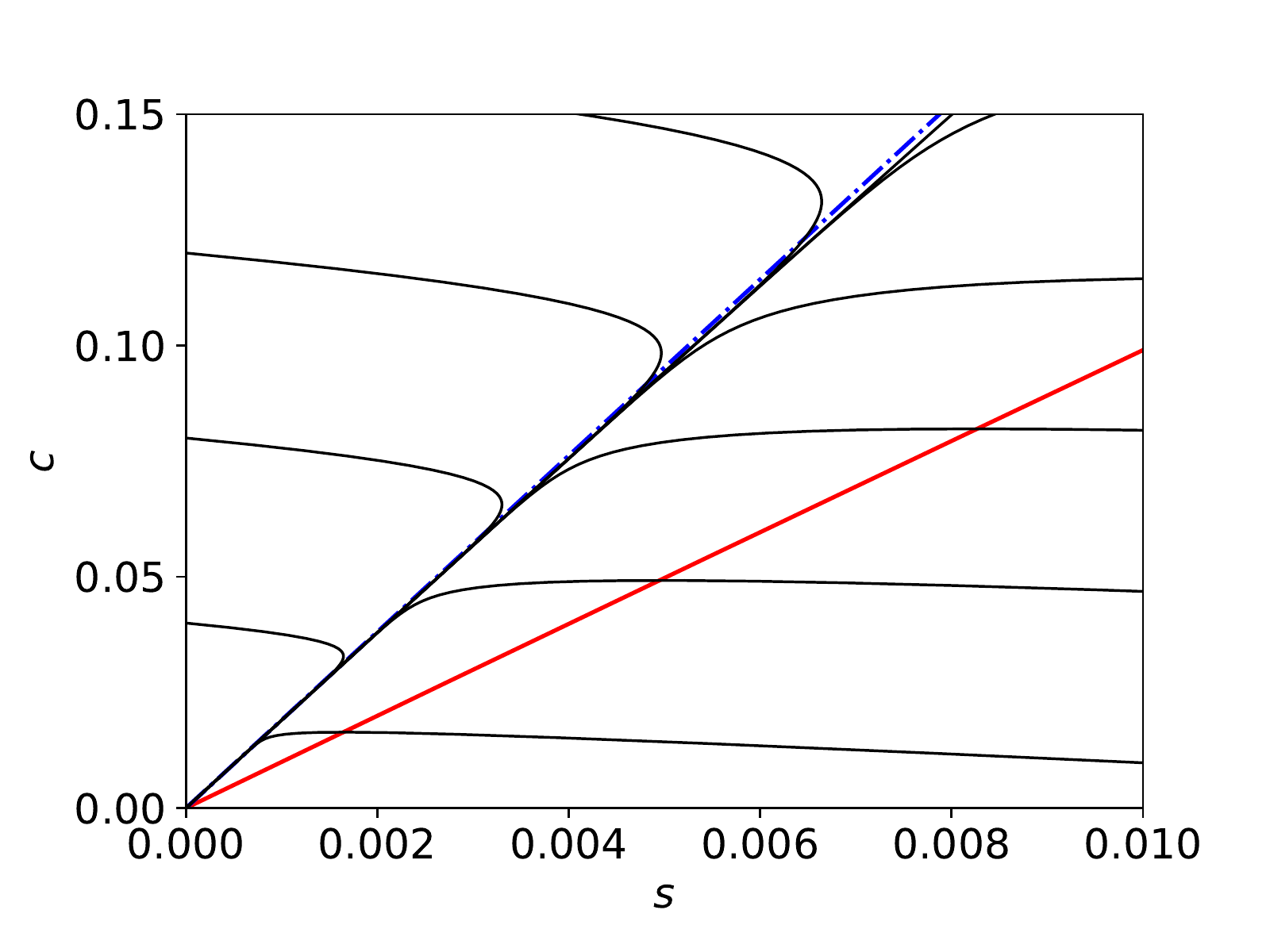}
    \includegraphics[width=8.0cm]{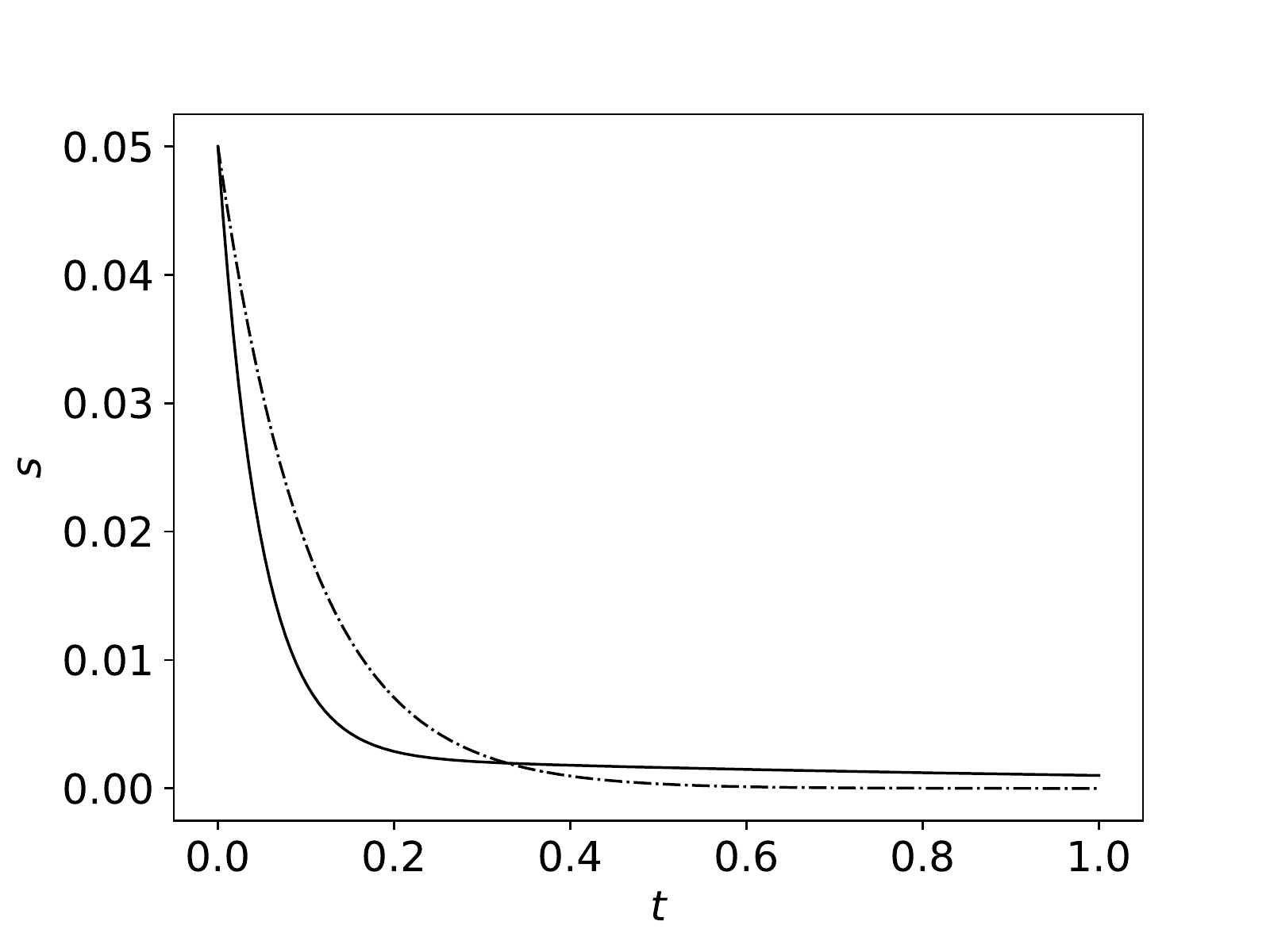}
  \caption{\textbf{The tangency condition must hold if phase plane trajectories are to approach 
  the QSS manifold towards the origin in the linear regime.} {{\sc Left}}: The thin black curves 
  are numerical solutions to the mass action equations~(\ref{eqmmirrev}) for the Michaelis--Menten 
  reaction mechanism. The red curve is the   QSS manifold corresponding to the $c$-nullcline. The 
  dashed/dotted blue line is the slow   eigenspace. Parameter values: $k_1=2.0$, $k_2=1.0$, $e_0=10.0$ 
  and $k_{-1}=1.0$ and   {$\delta {\approx} 0.0413$.} Eventually, trajectories 
  approach the origin in a direction that 
  is tangent to the slow eigenspace. However, since the QSS manifold is not tangent to the 
  slow eigenspace, the sQSSA must fail in the long-time limit. {{\sc Right}}: The solid black 
  curve is the numerical solution to the mass action equations~(\ref{eqmmirrev}). The dashed/dotted black 
  curve is the numerical solution to the sQSSA. In this example, the QSS manifold is far 
  from tangent to the slow eigenspace. This translates to an erroneous timescale estimate 
  for the degradation of $s$.} \label{fig:tan}
\end{figure}

At this juncture, it is imperative to recall the precise nature of the tangency condition. As 
an example, consider the following parameter values: $k_1=1.0$, $e_0=10.0$, $k_{-1}=10.0$ and 
$k_2=0.01$. It is straightforward to verify that
\begin{equation*}
    |g'(0)-\alpha| {\approx} 5 \times 10^{-4},
\end{equation*}
and therefore the QSS variety is practically tangent to the slow eigenspace at the origin. 
However, the sQSSA fails to capture the correct time evolution by a factor
\begin{equation*}
  \exp(\mu)=\exp\left(k_{-1}\left|g^\prime(0)-\alpha\right|/|\lambda_1|\right){\approx} 2.74.  
\end{equation*}

\begin{figure}
  \centering
    \includegraphics[width=8.0cm]{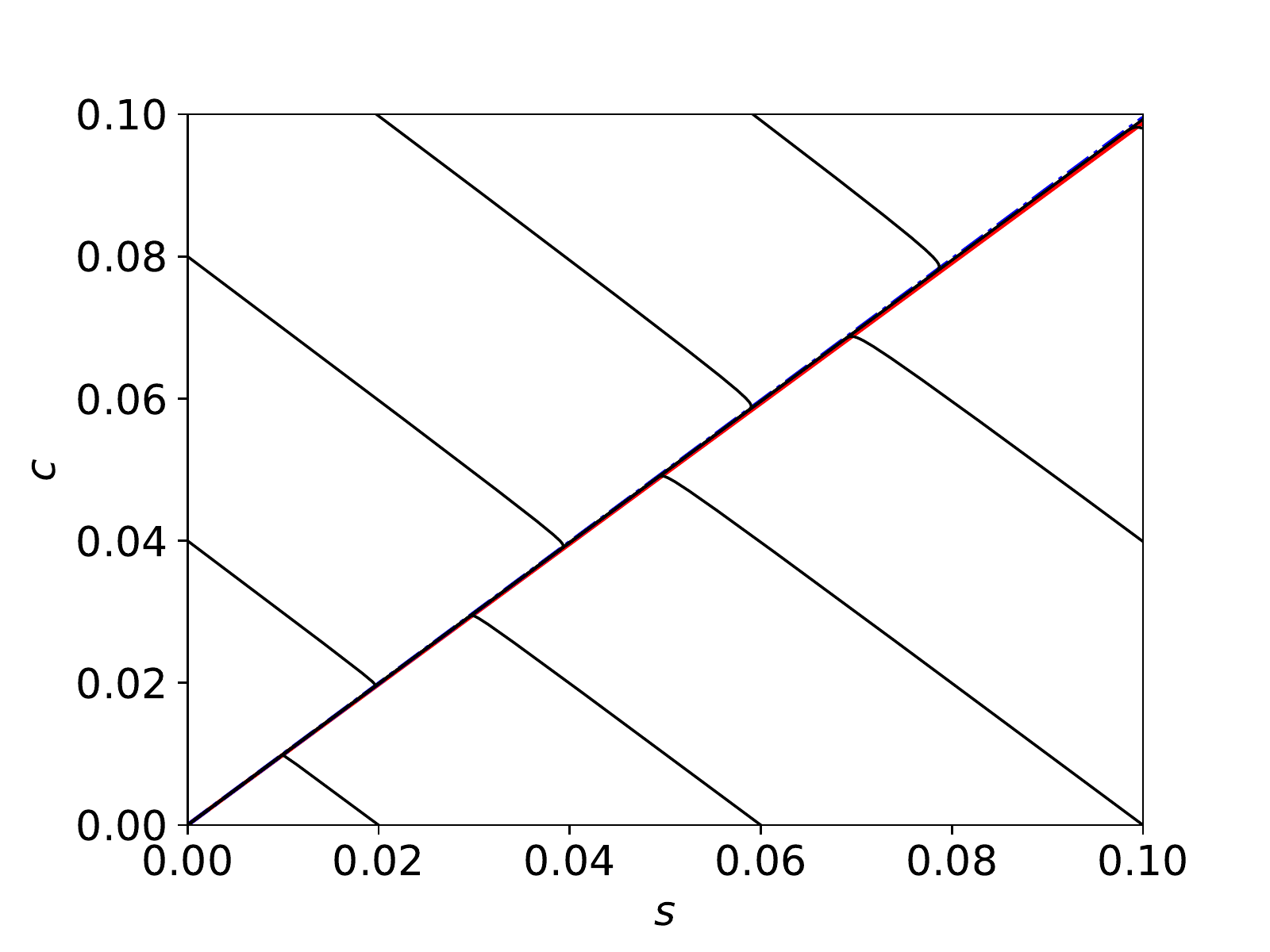}
    \includegraphics[width=8.0cm]{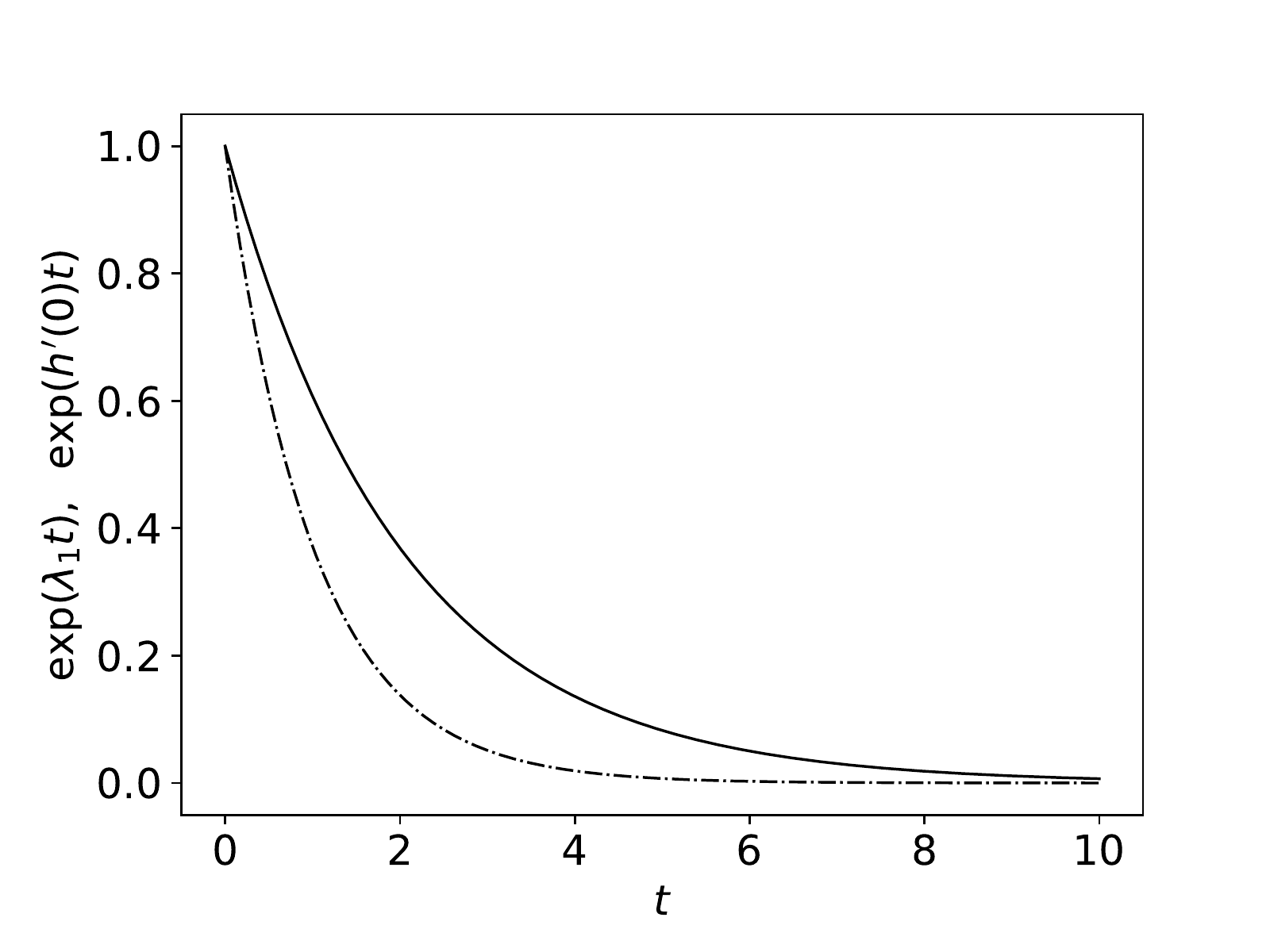}
  \caption{\textbf{Considering only the phase portrait may be deceptive: ``Geometric tangency'' 
  of the QSS variety to the slow eigenspace at the origin by itself does not guarantee the 
  long-time validity of the sQSSA.} {{\sc Left}}: The solid black curves are the numerical 
  solutions to the mass action equations ~(\ref{eqmmirrev}) for the Michaelis--Menten 
  reaction mechanism. The solid red line is the QSS variety, and the broken blue line is the 
  slow eigenspace. Note that these are virtually indistinguishable. Clearly, the phase-plane 
  trajectories approach the origin in the direction of the slow eigenspace, which is practically 
  tangent the QSS variety given by the $c$-nullcline. 
  {{\sc Right}}: The sQSS (dashed/dotted black curve) significantly underestimates the characteristic 
  timescale of substrate depletion in the linear regime. The true linear depletion rate is given 
  by the solid black curve. In both panels: $k_1=1.0$, $e_0=10.0$, $k_{-1}=10.0$, and $k_2=0.01$.} \label{fig:tanII}
\end{figure}

\begin{figure}
  \centering
    \includegraphics[width=8.0cm]{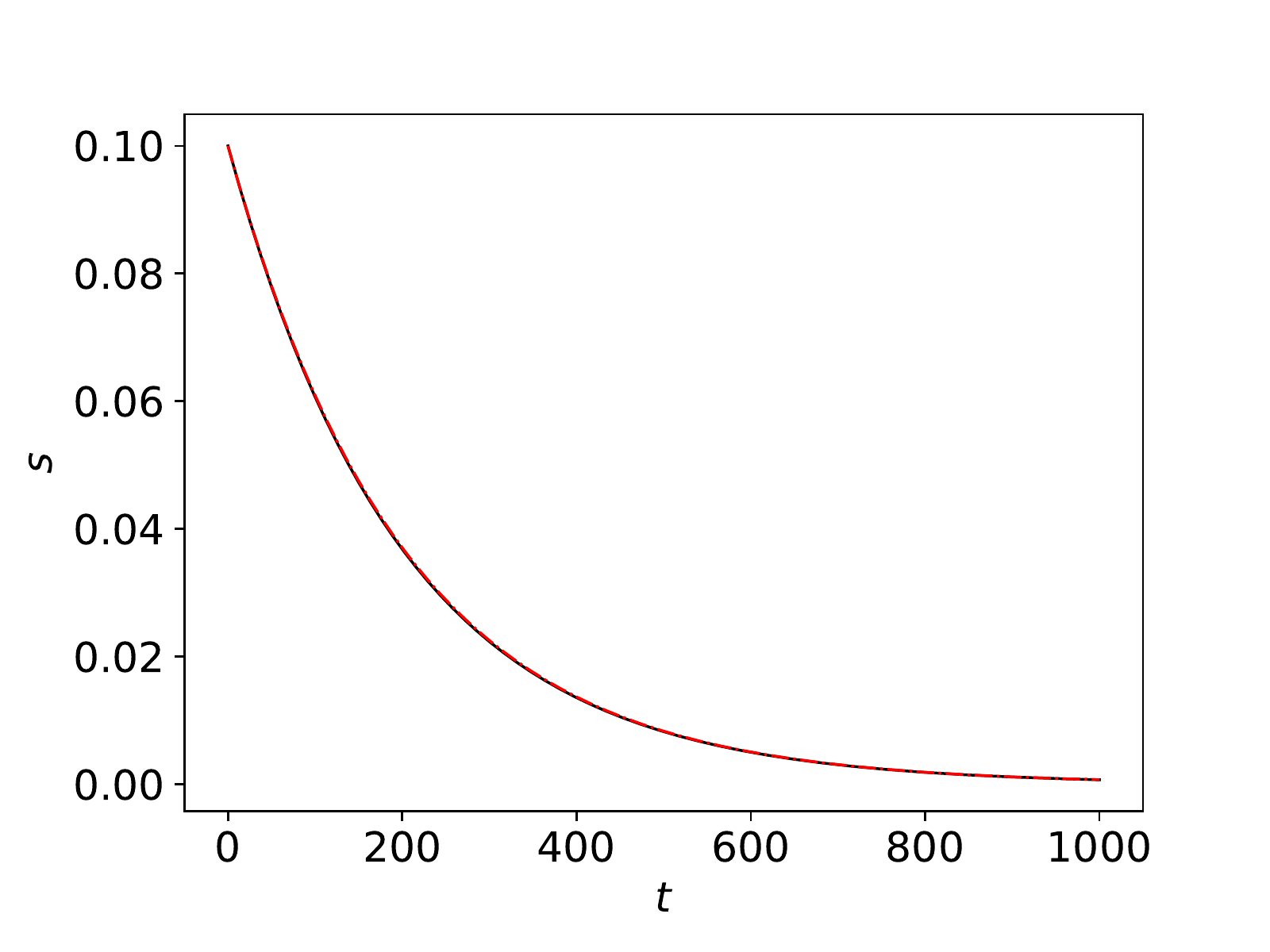}
    \includegraphics[width=8.0cm]{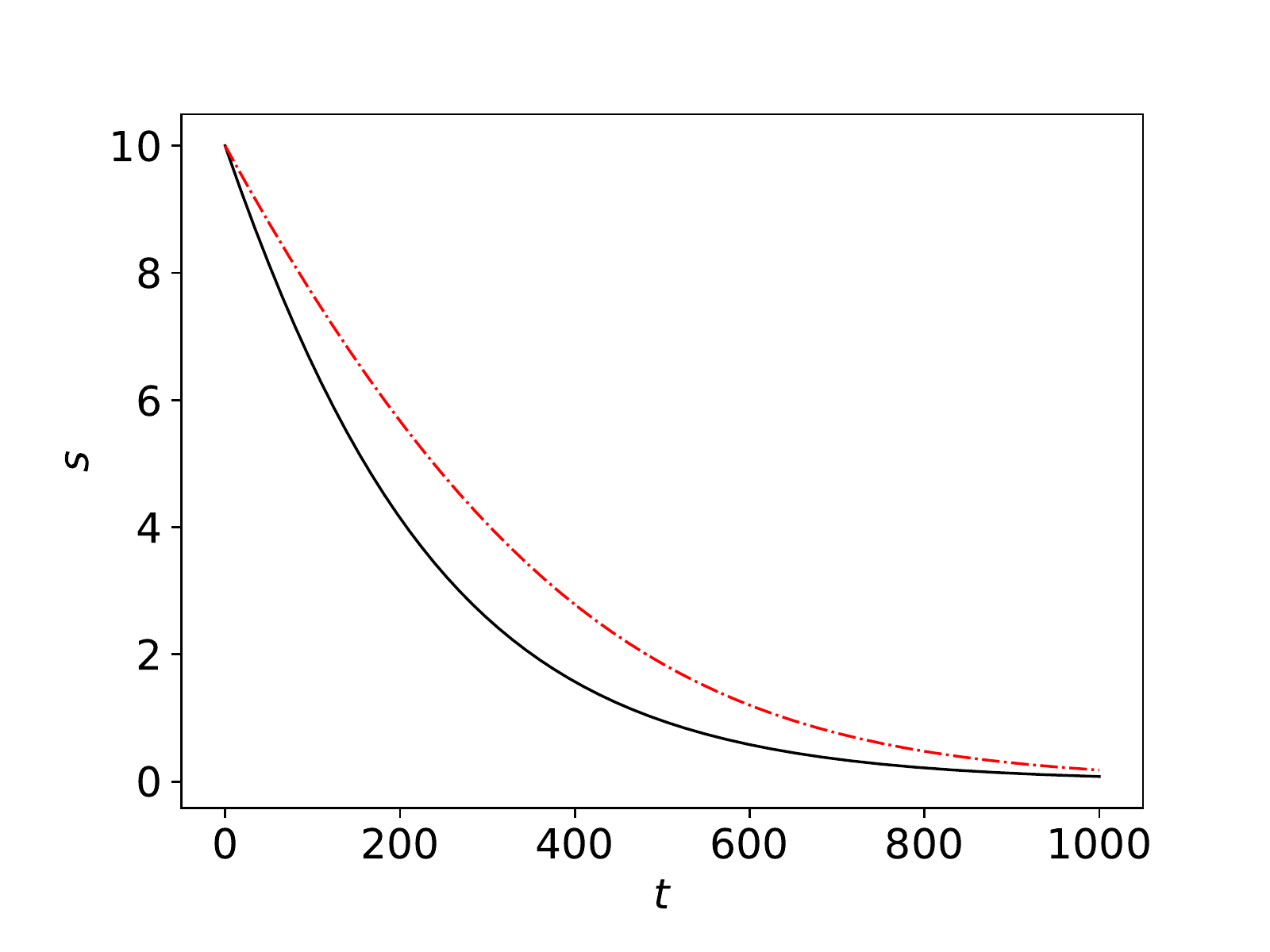}
  \caption{\textbf{Local validity of a QSS reduction does not imply global validity.} 
  {{\sc Left}}: In both panels, the solid black curve represents the numerical solution to the mass 
  action equations~(\ref{eqmmirrev}) for the Michaelis--Menten reaction mechanism; the dashed/dotted 
  red curve is the numerical solution to (\ref{globalCS}). Parameters are $k_1=1.0$, $e_0=10.0$, 
  $k_{-1}=10.0$, and $k_2=0.01$. {{\sc left}}: The initial condition used in the simulation is 
  $(s,c)(0)\approx (0.1, 0.01)$, and it is clear that (\ref{globalCS}) is a very good approximation 
  in the linear regime. {{\sc Right}}: The initial condition is $(s,c)(0)\approx (10,5)$ and one sees 
  that, outside of the linear regime, the reduced equation (\ref{globalCS}) (dashed/dotted red curve) 
  dramatically underestimates the rate of substrate depletion. The true linear depletion rate is given 
  by the solid black curve.} \label{fig:tanIII}
\end{figure}

{This example illustrates  that it is not sufficient to consider only the 
``geometric tangency'' of trajectories,\footnote{By \textit{geometric tangency} we are referring to 
{smallness of the term} $|g'(0)-\alpha|$ in (\ref{eq:mu}).} as is common practice 
in the literature on QSSA. The failure of the sQSS in {{\sc figure}} (\ref{fig:tanII}) is due to 
the presence of the ``amplification term'', $k_{-1}/|\lambda_1|$, in (\ref{eq:mu}). A sufficiently 
large $k_{-1}/|\lambda_1|$ can amplify a small difference between $g'(0)$ and $\alpha$. The error 
observed in (\ref{fig:tanII}) can be reduced by replacing $c=g(s)$ with an appropriate QSS manifold 
that improves the geometric tangency. For example, the $\sigma$--isocline of Calder and Siegel~\cite{CaSi}  
approaches the origin exactly tangent to the slow subspace. In dimensional form, the 
$\sigma$--isocline, $\sigma(s)$, is}
{
\begin{equation}
    c=\cfrac{\alpha e_0s}{e_0 + \alpha s} =:\sigma(s).
\end{equation}}
{It is straightforward to verify that $\sigma'(s)=\alpha$, and thus $\mu$ is identically 
zero. Utilizing the $\sigma$--isocline as our QSS variety yields}
{
\begin{equation}\label{globalCS}
    \dot{s} -\cfrac{e_0s}{e_0+\alpha s}\cdot (k_1e_0+k_{-1}\alpha),
\end{equation}}
{the linear approximation of which is exactly $\dot{s}=-\lambda_1s$. Consequently, 
the reduction (\ref{globalCS}) holds nicely near the stationary point; see 
{{\sc figure}}~\ref{fig:tanIII}, {{\sc left}} panel. However, the local validity of (\ref{globalCS}) 
does not guarantee its validity in the nonlinear regime; see 
{{\sc figure}}~\ref{fig:tanIII}, {{\sc right}} panel. Hence, local validity does not imply global 
validity in general.}
\section{Short-term versus long-term validity of QSSA}
Our results indicate that even for arbitrarily small Heineken et al.~\cite{hta} parameter
$\varepsilon_{HTA}$ or arbitrarily small Segel and Slemrod~\cite{SSl} parameter $\varepsilon_{SSl}$ 
the QSSA is not necessarily globally accurate. In fact, the initial substrate concentration 
is of no relevance for long-term QSS.\footnote{{This observation has been made before. 
Please read also, {\it Comment on the criterion $e_0 \ll s_0$}, Chapter~5, page~84 of 
Palsson~\cite{PalBook}, and see Patsatzis and Goussis \cite{PaGo}, in particular Fig. 3.}} This 
stands in stark contrast to the widely held belief by practitioners that $e_0\ll s_0$ is the 
relevant condition for QSSA. 

\begin{figure}
    \centering
    \includegraphics[scale=0.50]{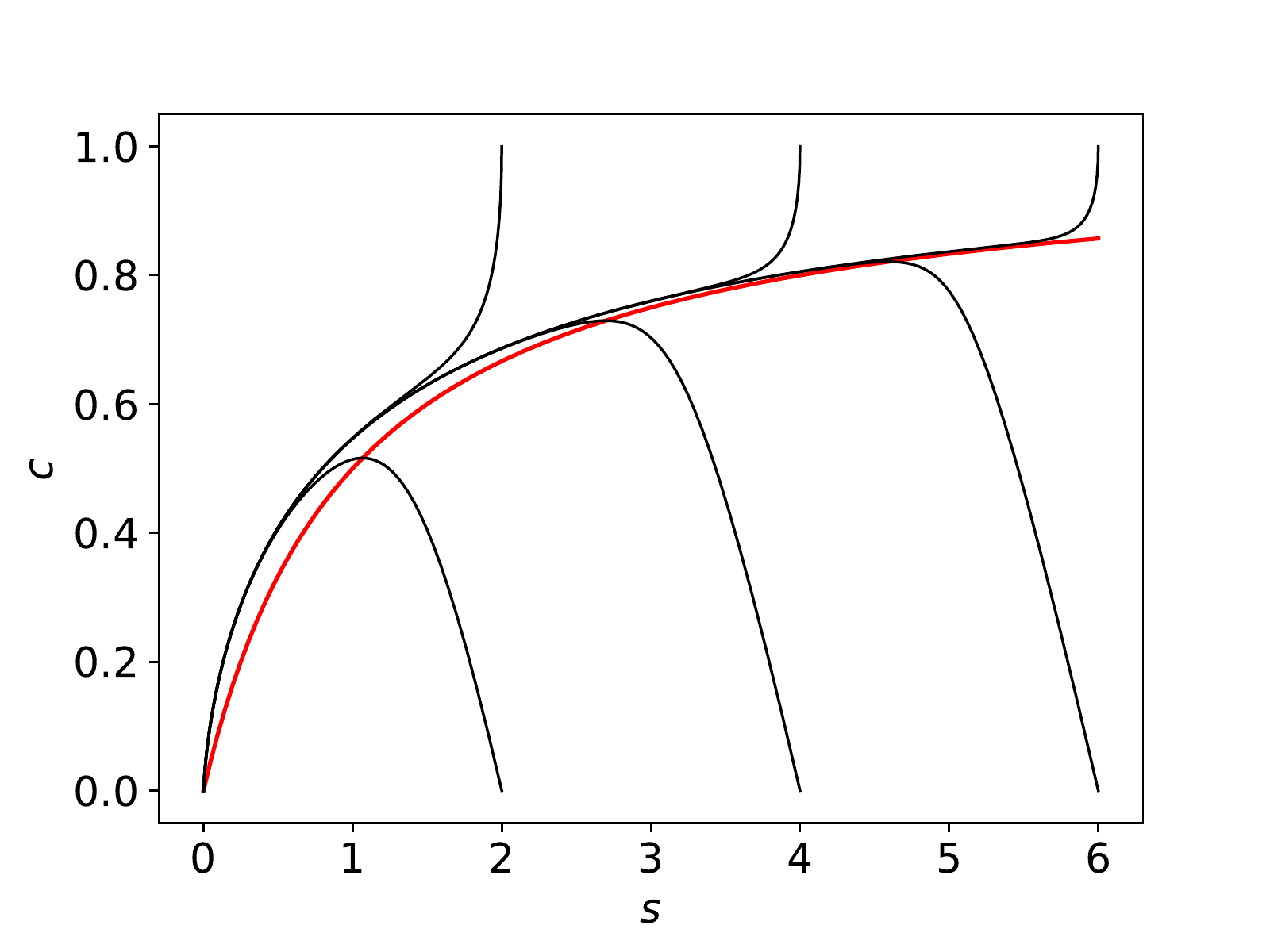}
    \includegraphics[scale=0.50]{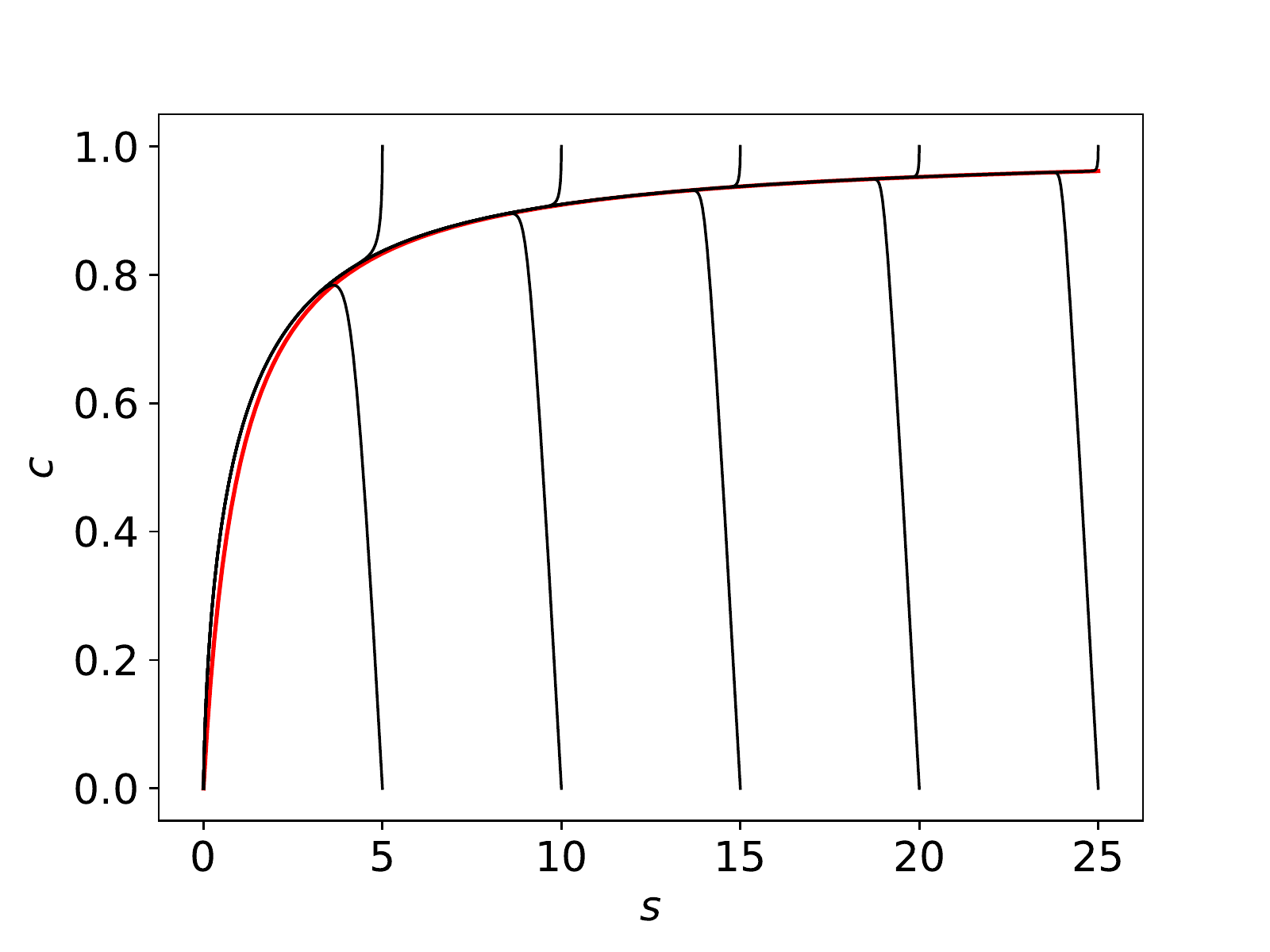}
    \caption{\textbf{The sQSSA is valid while $s$ is sufficiently large, even in the absence 
    of a spectral gap}. The solid black curves are numerical solutions to the mass action 
    equations~(\ref{eqmmirrev}) for the Michaelis--Menten reaction mechanism. The solid red 
    curve is the QSS manifold (i.e., the $c$-nullcline), given by $c=e_0s/(K_{\rm M}+s)$. The 
    parameters employed in the simulation are: $k_1=e_0=k_2=1.0$, $s_0=25.0$, and $k_{-1}=0$.
    Thus, $\delta=1$. Observe that even though the eigenvalues of the Jacobian at the origin 
    are equal and therefore the long-time validity of the QSSA fails ({{\sc Left}}), the 
    sQSSA still holds for sufficiently large values of $s$, which is illustrated in the 
    {{\sc Right}}. 
    \label{fig:traj}}
\end{figure}

We will at least partly reconcile these perspectives. In the limit $e_0\to 0$, with $s_0$ bounded, 
QSS reduction works indeed by singular perturbation theory, but QSS with increasing $s_0$ cannot 
be traced back to Tikhonov-Fenichel. However, one can actually show that in regions with sufficiently 
high concentrations of substrate -- corresponding to high $s_0$ and an early phase of the reaction -- 
the QSS condition holds with good accuracy. For an illustration, see {\sc{Figure}}~\ref{fig:traj}. 

\begin{figure}
    \centering
    \includegraphics[scale=0.5]{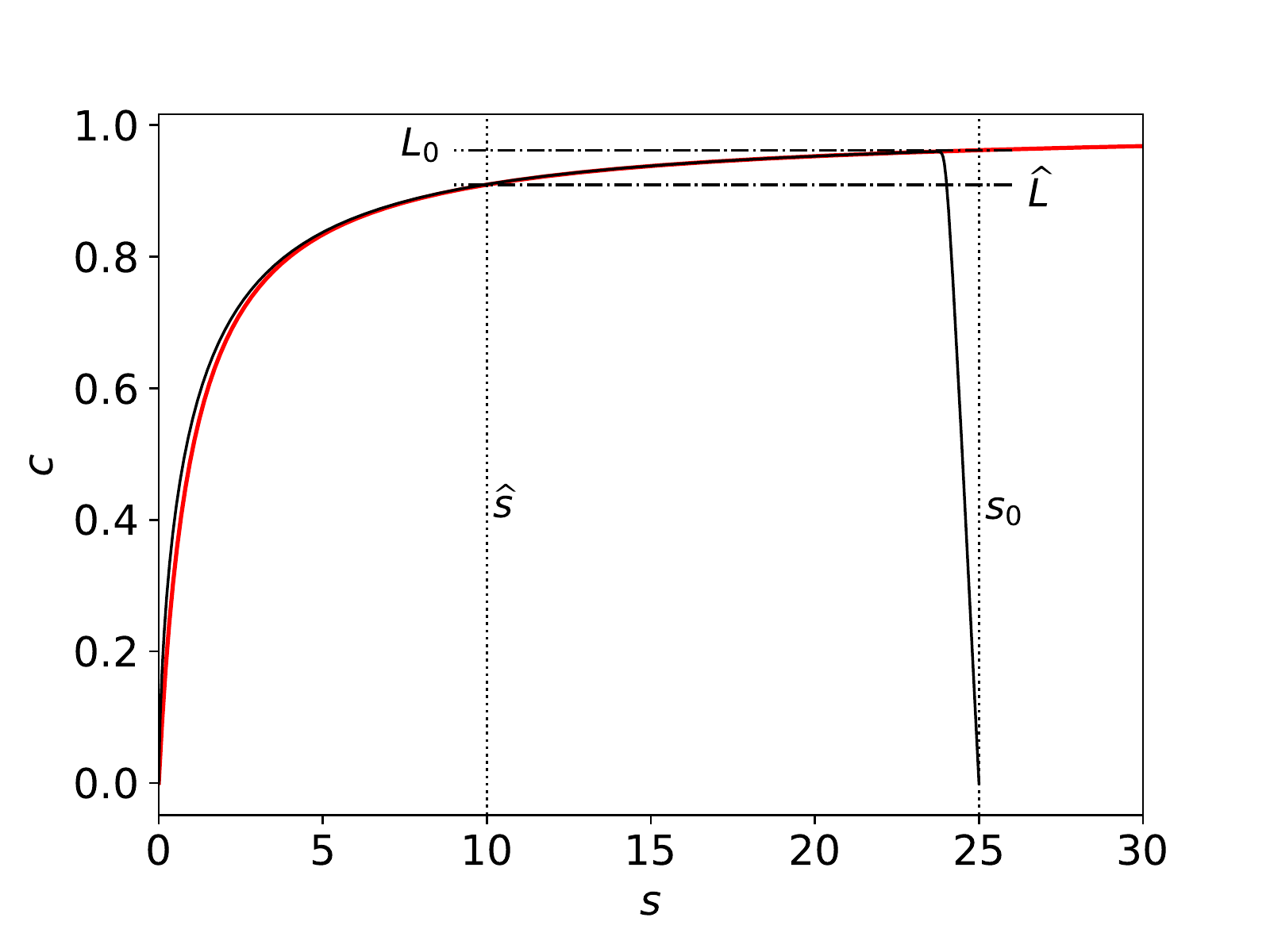}
    \includegraphics[scale=0.5]{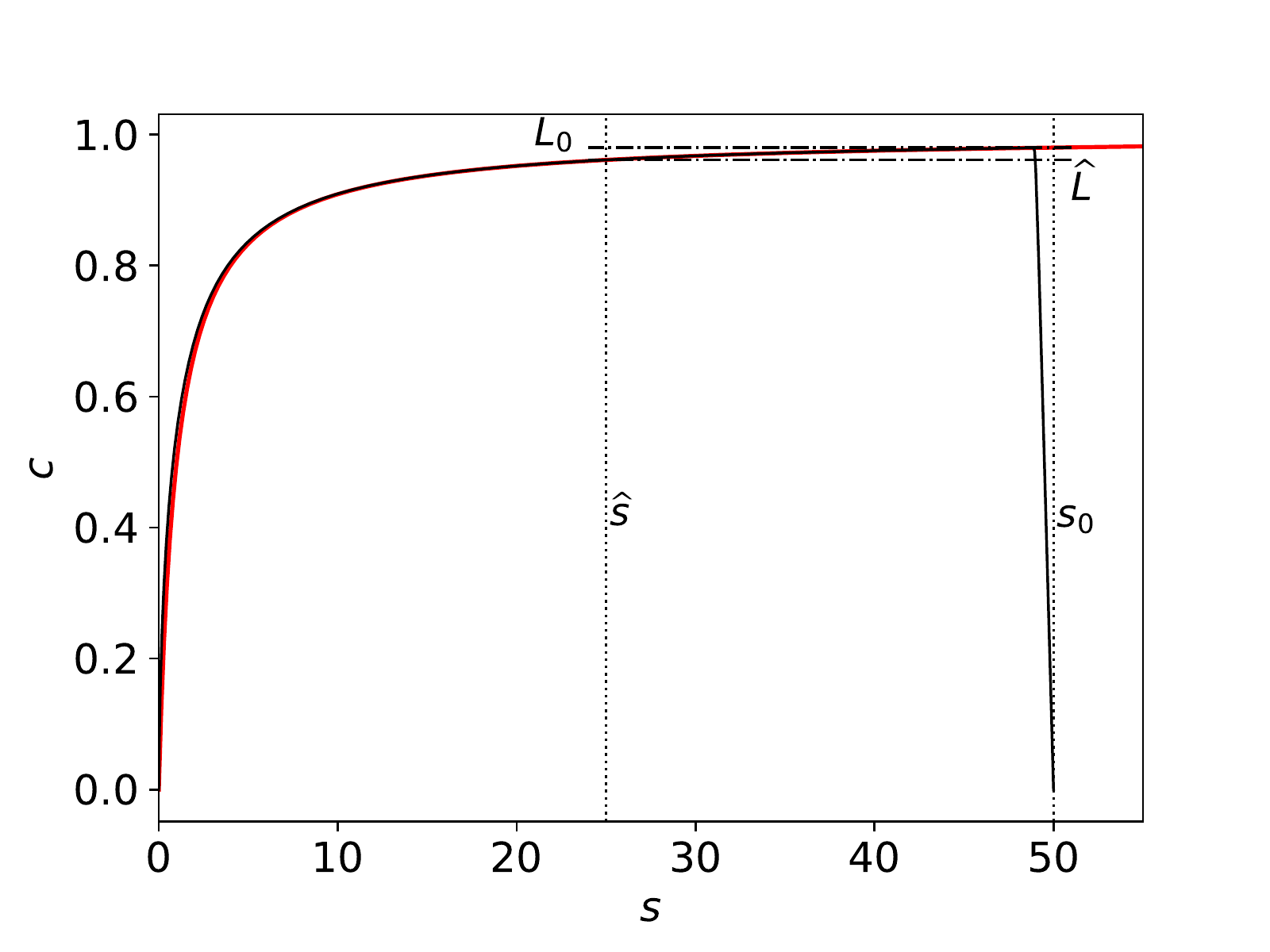}
    \caption{\textbf{The QSS is valid in domains where $s\geq \widehat s$, with $K_{\rm M}/\widehat{s}$ 
    sufficiently small}. In both numerical simulations, the solid black curve is the numerical 
    solution to the mass action equations~(\ref{eqmmirrev}) for the Michaelis--Menten reaction 
    mechanism. The solid red curve is the QSS manifold (i.e., the $c$-nullcline), given by 
    $c=e_0s/(K_{\rm M}+s)$. The dashed/dotted lines correspond to the limits $L_0$ and $\widehat{L}$. 
    The locations of $s_0$ and $\widehat{s}$ are demarcated by dotted vertical lines. 
    {{\sc Left}}: The parameters employed in the simulation are: $k_1=e_0=k_2=1.0$, 
    $s_0=25.0$, $k_{-1}=0$ $s_0=25$ and $\widehat{s}=10$. Observe that, after a brief transient, 
    the trajectory closely follows the QSS manifold for $s\in [\widehat{s},s_0]$, but departs 
    from the QSS manifold as it approaches the origin. {{{\sc Right}}}: The parameters employed 
    in the simulation are: $k_1=e_0=k_2=1.0$, $s_0=25.0$, $k_{-1}=0$ $s_0=50$ and $\widehat{s}=25$. 
    Observe that the distance between the upper and lower limits diminishes as 
    $\widehat s/K_{\rm M}$ increases.}
    \label{fig:traj2}
\end{figure}

To state our assertion precisely, let us consider a trajectory that starts at $(s_0,0)$ while 
it is confined to the region
\begin{equation*}
    \Lambda:=\{(s,c)\in \mathbb{R}^2_{\geq 0}: 0 \leq c \leq e_0\;\;\text{and}\;\; \widehat{s} \leq s \leq s_0\},\quad \widehat s\geq 0.
\end{equation*}
While $c\leq e_0$ holds throughout, $\Lambda$ is not positively invariant whenever $\widehat{s}>0$. 
From \eqref{eqmmirrev} we have
\begin{equation}\label{cEQ}
\dot c= k_1(e_0-c) s-(k_{-1}+k_2)c,
\end{equation}
and with $e_0-c\geq 0$ we obtain the 
differential inequalities
\begin{subequations}\label{INEQ1}
\begin{align}
\dot c&\geq k_1(e_0-c)\widehat s-(k_{-1}+k_2)c,\\
\dot c&\leq k_1(e_0-c)s_0-(k_{-1}+k_2)c.
\end{align}
\end{subequations}
With $c(0)=0$, equation (\ref{INEQ1}) implies
\begin{equation}\label{lowerestc}
c\geq\dfrac{e_0\widehat s}{K_{\rm M}+\widehat s}\left(1-\exp\left(-\widehat{\lambda}t\right)\right),\quad \text{for all } t \text{ such that }(s,c)\in \Lambda,
\end{equation}
by standard theorems on differential inequalities, where $\widehat{\lambda}:=k_1\widehat{s}+k_{-1}+k_2$.
Likewise, it also holds that
\begin{equation}\label{upperestc}
    c\leq\dfrac{e_0s_0}{K_{\rm M}+s_0}\left(1-\exp\left(-\lambda_0t\right)\right), \quad \text{ for all } t \geq 0,
\end{equation}
where $\lambda_0:=k_{-1}+k_2+k_1s_0$. Finally, using \eqref{eqmmirrev} with ``frozen'' 
$s\in[\widehat s, s_0]$ and invoking differential inequalities again, we arrive at
\begin{equation}
 \widehat L:=   \dfrac{e_0\widehat s}{K_{\rm M}+\widehat s}\leq \dfrac{e_0 s}{K_{\rm M}+s}\leq \dfrac{e_0s_0}{K_{\rm M}+s_0}=:L_0
\end{equation}
by taking the limit $t\to\infty$. Moreover, for \textit{large} $\widehat{\lambda}$  we see that 
the terms on the right hand sides of (\ref{lowerestc}) and (\ref{upperestc}) quickly approach 
their respective limits, $\widehat{L}$ and $L_0$. From Proposition~2 by Noethen and Walcher~\cite{NoWa} 
(see also Calder and Siegel~\cite{CaSi}), it is known that the trajectory starting 
at $(s_0,0)$ crosses the QSS manifold $Y$. From Proposition~12 by Noethen and Walcher~\cite{NoWa}, 
after a characteristic time that is ${O}(1/|\widehat{\lambda}|)$, it holds that 
$\widehat{L} \leq c \leq L_0.$ Furthermore, the distance between $\widehat{L}$ and $L_0$ is given by
{\begin{equation*}
    L_0-\widehat{L}=\cfrac{K_{\rm M}(s_0-\widehat s)e_0}{(K_{\rm M}+\hat{s})(K_{\rm M}+s_0)}\ ,
\end{equation*}}
which is practically negligible whenever $K_{\rm M} \ll \widehat{s}$ and $e_0$ bounded. For 
instance, with $\widehat s=s_0/2$ one has
{\begin{equation*}
    L_0-\widehat{L}=L_0\cdot\cfrac{K_{\rm M}}{2(K_{\rm M}+\widehat{s})}.
\end{equation*}}
Consequently, the phase plane trajectory will remain very close to the QSS manifold everywhere 
in $\Lambda$ where $K_{\rm M}\ll \widehat{s}$. In this sense, the QSS for complex is 
valid. For a numerical validation, please see {{\sc Figure}}~\ref{fig:traj2}.

It is clear from {\sc Figure}~\ref{fig:traj2} that large initial substrate concentration will 
ensure the temporary validity of the QSSA but, without a sufficient spectral gap, the sQSSA 
will lose accuracy once $s \lesssim K_{\rm M}$. One can also invoke the following argument to 
understand that increasing $s_0$ cannot improve the quality of the QSS reduction over the full 
course of the reduction: Moving a point along a trajectory just amounts to translation of 
time, which leaves the long-term behavior unaffected.

Moreover, this loss may be consequential in the context of progress curve experiments utilized 
to estimate $K_{\rm M}$ and the maximal turnover rate $V=k_2e_0$ by fitting time-course data to the sQSSA. 
According to Stroberg and Schnell~\cite{Strob}, measurements should be taken with $s$ near 
$K_{\rm M}$, as the inverse problem is ill-posed in regions where the QSS variety is extremely flat, 
i.e., where $K_{\rm M} \ll s$. The questionable quality of parameter fitting also turns up in the 
initial rate experiments with varying, but high initial substrate concentrations $s_0$, since all 
the data points for regression to determine the Lineweaver-Burk line lie close to the ordinate, 
and thus proper extrapolation becomes doubtful. The same challenge will apply to initial rate
experiments carrying out parameter estimation to a non-linear fitting of the Michaelis--Menten
equation~\eqref{smmred}. Consequently, an overabundant concentration of substrate may not be 
strategically beneficial for parameter estimation applications, and thus we see that the local
validity of the QSSA has experimental (in addition to merely mathematical) relevance. 
{The inverse modeling problem of parameter estimation needs to be investigated
systematically based on the findings of this work.}

\section{Discussion}
In Section \ref{loccsec} the validity of local QSSA for the Michaelis--Menten reaction mechanism has been 
reconsidered, and has been shown to be fully compatible with the fundamental requirement of eigenvalue disparity. The eigenvalue disparity conditions are the same as those given by Patsatzis and Goussis \cite{PaGo}, specialized to the stationary point\footnote{Patsatzis and Goussis emphasize employing the eigenvalue ratio condition globally, following established practice from computational singular perturbation (CSP) theory. Thus they obtain a candidate for a global invariant manifold. The existence of a global slow manifold is known from Calder and Siegel \cite{CaSi}; see also Eilertsen et al.\ \cite{ERSW}. It would seem interesting to discuss how these global results relate to each other.}, and they are also present in earlier work by Palsson and Lightfoot \cite{PaLi}. But (implicitly) in these works the eigenvalue conditions have only been considered with sufficiency in mind.\footnote{A similar remark seems to apply to the derivation of several types of small parameters in the literature.} From a mathematical perspective, in order to prove that the conditions are also necessary, one has to establish that QSSA actually fails when they are violated. This is the new aspect in the present work, where we have determined, quantitatively, a region in parameter space where
any type of QSS fails to hold (anti-QSS) due to the absence of a sufficiently large spectral gap. 
The fact that a random choice of reaction parameters will likely lie outside this region, may explain 
the widely held belief that some type of QSSA (even more specifically, tQSSA) is roughly valid 
for all parameters. On the other hand, the analysis by Patsatzis and Goussis~\cite{PaGo} 
shows conclusively that some type of QSSA does hold in large regions of parameter space.

Furthermore, in Section \ref{ltsubsec} we have shown that, for any given QSSA, the approximate tangency of the QSS variety 
to the slow eigenspace of the stationary point (characterized by $\mu \ll 1$, with $\mu$ given 
in \eqref{eq:mu}) is necessary to ensure good approximation by the QSS reduction. For the 
sQSSA and tQSSA, we have shown that the Reich and Selkov~\cite{ReSe} condition, $e_0 \ll K_{\rm M}$, 
ensures the long-time validity of the sQSSA, as this condition ensures the existence 
of a spectral gap as well as the tangency condition, $\mu \ll 1$.  Note that this condition
does not involve initial substrate.

We address the difference between the commonly used parameters due to Heineken et al.~\cite{hta},
{respectively} to Segel and Slemrod~\cite{SSl}, and the ones derived in the 
present paper. The discrepancy is due to Segel's choice of a nonlinear timescale estimate, 
and we showed that it may provide incorrect predictions for the existence and the nature of 
a reduction over the whole time course. In a way, this is not unexpected. Initial conditions 
can be moved along trajectories, and this does not influence the long-term behavior.

Our analysis has resulted in a considerably clearer understanding of the QSSA, and also 
our discussion points the way to an improvement in a general technique of applied
mathematics. There are important lessons to be drawn from the anti-QSSA. A number of
papers in the literature assume that certain QSS reductions will always be roughly valid 
across the full parameter space. Our results show that this is not the case. More
importantly, the conditions for the anti-QSSA presents an important opportunity to 
assert the region of the parameter domain where no QSSA can be applied to model systems 
or use QSS reduction to estimate parameters. This is an important subject of 
active research in rigor and reproducibility~\cite{Hall,Strob}. 

The moral of this paper seems to be that estimating the validity of QSS reductions
is not as \textit{mechanical} as it appears. For planar systems, any qualifier (i.e., dimensionless parameter) claiming to certify the legitimacy of a QSS reduction must, at minimum, satisfy the requirements discussed in this work. Based on our survey of the QSS literature, it appears that intuitive scaling arguments sometimes produce qualifiers that fail to meet these obligations. The fundamental requirements highlighted in this work should be considered in future applications if progress in deterministic QSS theory (and possibly stochastic QSS theory) is to advance. Moreover, the extensive utilization of QSS reduction in mathematical biology and applied mathematics suggests that additional progress is indispensable.

\section{Appendix}\label{app}
Here we provide some details on the function $q$ introduced in Section~\ref{ltsubsec}, and 
proceed to obtain sharper estimates on the parameter $\mu$. 

Differentiating the expression in \eqref{qfunc}, one finds
\[
q'(x)=-\frac12+\frac{k_2}{k_{-1}+k_2}+\frac12\cdot \dfrac{x+k_{-1}-k_2}{\left((x+k_{-1}+k_2)^2-4k_2x\right)^{1/2}}
\]
and furthermore
\begin{equation}\label{q2prime}
    q''(x)=\dfrac{2k_{-1}k_2}{\left((x+k_{-1}+k_2)^2-4k_2x\right)^{3/2}}> 0\text{ for }x>0,
\end{equation}
which implies all the assertions in Section~\ref{ltsubsec}.

More detailed estimates start from observing
\begin{equation}
    q(x)= q''(\xi)\cdot x^2\text{  for some  }\xi,\quad 0\leq \xi\leq x,
\end{equation}
by Taylor's theorem, and from upper and lower estimates for the maximum and minimum of 
$q''(\xi)$ on the interval $[0,x]$. As evident from \eqref{q2prime}, this maximum and 
minimum correspond to the minimum and maximum (in this order) of the quadratic function
\begin{equation}
\eta(\xi):=(\xi+k_{-1}+k_2)^2-4k_2\xi=(\xi+k_{-1}-k_2)^2+4k_{-1}k_2 \text{ on } [0,x].
\end{equation}
One has to distinguish cases now:
\begin{itemize}
    \item In case $k_{-1}\geq k_2$, the minimum appears at $\xi=0$, with value 
    $(k_{-1}+k_2)^2$, and the maximum appears at $\xi=x$.
    \item In case $k_{-1}<k_2$ and $x\leq 2(k_2-k_{-1})$, the minimum appears at 
    $\xi=k_2-k_{-1}$, with value $4k_{-1}k_2$, and the maximum appears at $\xi=0$.
    \item In case $k_{-1}<k_2$ and $x> 2(k_2-k_{-1})$, the minimum appears at 
    $\xi=k_2-k_{-1}$, with value $4k_{-1}k_2$, and the maximum appears at $\xi=x$.
\end{itemize}
Thus, for instance, in case $k_{-1}\geq k_2$ the maximum of $q''$ appears at $\xi=0$, 
with value $2k_{-1}k_2/(k_{-1}+k_2)^3$, and one gets 
\[
\lambda_1-h'(0)\leq \dfrac{2k_{-1}k_2\,(k_1e_0)^2}{(k_{-1}+k_2)^3}.
\]
Combining this with the estimate from \eqref{lamoneest} one arrives at
\begin{equation}\label{firstsharpest}
 \mu=   \dfrac{h'(0)}{\lambda_1}-1\leq \dfrac{2k_{-1}\cdot(k_{-1}+k_2+k_1e_0)\cdot k_1e_0}{(k_{-1}+k_2)^3}\leq \dfrac{2k_{-1}\cdot k_1e_0}{(k_{-1}+k_2)^2}.
\end{equation}
One can work through all the cases in a similar manner, with the following results:
\begin{itemize}
    \item In case $k_{-1}\geq k_2$, one has the upper estimates \eqref{firstsharpest} 
    for $\mu$, and the lower estimates
    \begin{equation}
        \mu\geq\dfrac{k_{-1}\cdot(k_{-1}+k_2+k_1e_0)\cdot k_1e_0}{\left((k_{-1}+k_2+k_1e_0)^2-4k_2k_1e_0\right)^{3/2}}\geq\dfrac{k_{-1}\cdot k_1e_0}{\left(k_{-1}+k_2+k_1e_0\right)^{2}}.
    \end{equation}
    \item In case $k_{-1}<k_2$ one gets the upper estimates
    \begin{equation}
         \mu\leq\dfrac{(k_{-1}+k_2+k_1e_0)\cdot k_1e_0}{4k_{-1}^{1/2}k_2^{3/2}}\leq\dfrac{(k_{-1}+k_2+k_1e_0)\cdot k_1e_0}{4k_{-1}^{2}}.
    \end{equation}
    Furthermore:
    \begin{itemize}
        \item In case $k_1e_0\leq 2(k_2-k_{-1})$ one obtains the lower estimates
        \begin{equation}
            \mu\geq\dfrac{k_{-1}\cdot(k_{-1}+k_2+k_1e_0)\cdot k_1e_0}{(k_{-1}+k_2)^3}\geq\dfrac{k_{-1}\cdot k_1e_0}{(k_{-1}+k_2)^2}.
        \end{equation}
        \item In case $k_1e_0> 2(k_2-k_{-1})$ one obtains the lower estimate
        \begin{equation}
        \mu\geq\dfrac{k_{-1}\cdot(k_{-1}+k_2+k_1e_0)\cdot k_1e_0}{\left((k_{-1}+k_2+k_1e_0)^2-4k_2k_1e_0\right)^{3/2}}\geq\dfrac{k_{-1}\cdot k_1e_0}{\left(k_{-1}+k_2+k_1e_0\right)^{2}}.
        \end{equation}
    \end{itemize}
\end{itemize}
These estimates are more detailed than the ones presented in the main part of the 
paper, and in particular one obtains positive lower bounds for $\mu$ in every case. 
But the necessary distinction of cases may offset the advantage in applications.
Note that we do not aggregate the rate constants in terms of $K$ and $K_{\rm M}$ in this
Appendix.


\end{document}